\documentclass[graybox]{svmult}

\usepackage{mathptmx}       
\usepackage{helvet}         
\usepackage{courier}        
\usepackage{type1cm}        
\usepackage{makeidx}         
\usepackage{graphicx}        
\usepackage{multicol}        
\usepackage[bottom]{footmisc}

\makeindex             

\usepackage{textcomp} 
\usepackage{amssymb}  
\usepackage{bm}       
\usepackage{booktabs} 
\usepackage{dcolumn}  
\usepackage{multirow} 
\usepackage[printonlyused]{acronym} 
\usepackage{amsmath, amsfonts}
\usepackage{float}
\usepackage{siunitx}
\graphicspath{{figures/}} 

\usepackage{pstricks}
\begin{document}

\title*{Proximity-induced topological transition and strain-induced charge transfer in graphene/MoS$_2$ bilayer heterostructures}

\author{Sobhit Singh, Abdulrhman M. Alsharari, Sergio E. Ulloa, and Aldo H. Romero}

\institute{Sobhit Singh and Aldo H. Romero \at Department of Physics and Astronomy, West Virginia University, Morgantown, West Virginia 26505, USA; E-mail: {\it{smsingh@mix.wvu.edu, Aldo.Romero@mail.wvu.edu}}  \and Abdulrhman M. Alsharari and Sergio E. Ulloa \at Department of Physics and Astronomy, Nanoscale and Quantum Phenomena Institute, Ohio University, Athens, Ohio 45701, USA; E-mail: {\it aalsharari@ut.edu.sa, ulloa@ohio.edu}}


\maketitle

\abstract{Graphene/MoS$_2$ heterostructures are formed by combining the nanosheets of graphene and monolayer MoS$_2$. The electronic features of both constituent monolayers are rather well-preserved in the resultant heterostructure due to the weak van der Waals interaction between the layers. However, the proximity of MoS$_2$ induces strong spin orbit coupling effect of strength $\sim$1 meV in graphene, which is nearly three orders of magnitude larger than the intrinsic spin orbit coupling of pristine graphene. This opens a bandgap in graphene and further causes anticrossings of the spin-nondegenerate bands near the Dirac point. Lattice incommensurate graphene/MoS$_2$ heterostructure exhibits interesting moir\'{e} patterns which have been observed in experiments. The electronic bandstructure of heterostructure is very sensitive to biaxial strain and interlayer twist. Although the Dirac cone of graphene remains intact and no charge-transfer between graphene and MoS$_2$ layers occurs at ambient conditions, a strain-induced charge-transfer can be realized in graphene/MoS$_2$ heterostructure. Application of a gate voltage reveals the occurrence of a topological phase transition in graphene/MoS$_2$ heterostructure. In this chapter, we discuss the crystal structure, interlayer effects, electronic structure, spin states, and effects due to strain and substrate proximity on the electronic properties of graphene/MoS$_2$ heterostructure. We further present an overview of the distinct topological quantum phases of graphene/MoS$_2$ heterostructure and review the recent advancements in this field. 
}

\section{Introduction}
\label{sec:intro}
The successful isolation of graphene from bulk graphite \cite{Novoselov2004} has triggered a new burgeoning research area in atomically thin two-dimensional (2D) materials. Since the last decade, several 2D materials namely - graphene, BN, MoS$_2$, MoSe$_2$, WS$_2$, WSe$_2$, MoTe$_2$, Xene sheets (X = Si, Ge, Sn), phosphorene, bismuthene, and many more, have been fabricated and extensively investigated due to their promising applications in the electronic, valleytronic, spintronic, catalysis, energy, and biosensing areas. \cite{graphene, geim2007rise, kis2012graphene, wang2012electronics, fiori2014electronics, balendhran2015elemental, schaibley2016valleytronics, C5NR01052G, molle2017buckled, C4CS00102H, SinghBiSb2017, NevalaitaPRB2018} Some of the notable properties that make 2D materials interesting are: high carrier mobility, superconductivity, mechanical flexibility, exceptional thermal conductivity, large photoluminescence, high optical and UV absorption, quantum spin Hall effect, strong light-matter interactions, and observation of highly confined plasmon-polaritons. \cite{ graphene, mak2012control, liu2015strong, low2017polaritons} Interestingly, these properties can be efficiently harnessed in 2D materials by means of strain engineering, number of atomic layers, adsorption, intercalation, interlayer twist, proximity effects and gate voltage. \cite{WooJong2012, Geim_vdw2013, Mishchenko2014, C4CS00287C} Furthermore, several types of 2D materials can be vertically stacked to design van der Waals (vdW) heterostructures which often enhance the desirable properties of the constituent atomic layers. \cite{WooJong2012, Geim_vdw2013, Mishchenko2014, jariwala2017mixed} These heterostructures offer unique ways to tailor their remarkable properties, hence they have promising applications in modern technology. However, control of the doping type, carrier concentration, and stoichiometry remains challenging in most of the known 2D materials and vdW heterostructures. \cite{jariwala2017mixed}

Graphene, a two dimensional monolayer of carbon atoms arranged in a honeycomb lattice, has emerged as the most celebrated 2D material of the last decade. It has been thoroughly investigated and many of its interesting features have been revealed \cite{graphene}. A single layer graphene exhibits numerous novel features such as ultra-high intrinsic mobility (200,000 cm$^{2}$/V$^{-1}$s$^{-1}$), large electrical conductivity, excellent thermal conductivity (5,000 W$^{-1}$K$^{-1}$), biosensing, and exceptional elastic and mechanical properties with a very large Young's modulus ($\sim$1.0 TPa). \cite{graphene, katsnelson2012graphene, aoki2013physics} However, the negligible intrinsic spin-orbit coupling (SOC) and correspondingly small energy bandgap limit many practical applications of pristine graphene in spintronics. In recent years, researchers have succeeded in enhancing the bandgap of graphene by several orders using unconventional methods and substrate proximity effects. The availability of many other 2D crystals allows us to design new graphene-based vdW heterostructures having strong proximity effects. A particular family of such 2D crystals is the semiconducting transition metal dichalcogenides (TMDs)- MX$_2$ (M = Mo, W and X = S, Se, Te) - that shows interesting optoelectronic and valleytronic features, and offer strong proximity effects on graphene's electronic band structure. \cite{mog,mogrf,w2,mogrf2,Abdulrhman2016}

Atomically thin MX$_2$ semiconductors (M=W, Mo and X= S, Se, Te) form a sandwich structure with a honeycomb lattice \cite{mos2}, where one atomic layer of transition-metal atom (M) is sandwiched between two atomic layers of chalcogens (X). These semiconductors exhibit a strong SOC in their valence bands, which increases with increasing mass of the M atom. MoS$_2$ is one of the most widely studied TMDs with a tunable bandgap in the visible and infrared (IR) regions of the electromagnetic spectrum as the number of atomic layers in the crystal changes. Bulk MoS$_2$ exhibits an indirect bandgap of $\sim$1.3 eV which increases with decreasing number of layers. \cite{wang2012electronics, Swastibrata2014, mog,mogrf} A monolayer of MoS$_2$ shows direct bandgap with energy gap of $\sim$1.8 eV at K \& K$'$ high symmetry points of the hexagonal Brillouin zone. Because of the broken inversion symmetry, SOC effects lift the spin-degeneracy of bands and substantially split the highest valence bands at the K \& K$'$ points. This broken spin degeneracy, when combined with the time-reversal symmetry present in pristine MoS$_2$, yields inherently coupled electronic bands at K \& K$'$ valleys which results in the possible observation of spin-valley effects and optical polarization memory in these materials. \cite{liu2015strong}

In pursuit of combining the novel features of graphene and MoS$_2$ monolayers, and mitigate their undesirable properties, researchers have recently made outstanding efforts to combine graphene and MoS$_2$ monolayers, and built graphene/MoS$_2$ vdW heterostructures. \cite{RoyNature2013,SimoneACSNano2013, SupNature2013, Geim_vdw2013, Mishchenko2014} Lattice incommensurate graphene/MoS$_2$ heterostructures show intriguing properties that can be controlled by tuning several factors such as strain, relative sliding between layers, interlayer twist, doping, bending, stacking order, and intercalation. \cite{BSachs_APL2013, Zilu2015, RMElder2015, Byungjin2015, Tribhuwan2016} Due to the lattice mismatch between graphene and MoS$_2$ monolayer, moir\'{e} patterns are expected to appear in graphene/MoS$_2$ vdW heterostructures, which has been observed in the recent experiments. \cite{G-MOS2-exp, w2, g-ws2} 

The proximity of MoS$_2$ induces relatively strong SOC effects in graphene opening an energy bandgap at the Dirac point. \cite{LuPRL2014} This bandgap can be further enhanced by means of gating and strain. Interestingly, the substrate induced SOC effects compete with the intrinsic SOC of graphene causing anti-crossing of spin-split bands near the Dirac point. \cite{Abdulrhman2016} One can also realize distinct topological quantum phases in graphene/MoS$_2$ heterostructures by exploiting an interlink between the proximity effects, SOC and the staggered potential. \cite{Abdulrhman2016} In a recent work, Gmitra et al. \cite{G-TMDs} have demonstrated that a SOC induced band-inversion occurs near the Dirac point in graphene/WS$_2$ heterostructure, thanks to the large SOC of W, which yields a quantum spin-Hall phase with chiral edge states in the graphene/WS$_2$ heterostructure. A similar topological phase transition can be realized in graphene/MoS$_2$ heterostructures by applying a gate voltage. \cite{Abdulrhman2016} In addition to these topological features, recent works report the observation of exceptional optical response with large quantum efficiency, gate-tunable persistent photoconductivity, excellent mechanical response, high power conversion efficiency, photocurrent generation, and negative compressibility in the graphene/MoS$_2$ heterostructures. \cite{RoyNature2013,SimoneACSNano2013, SupNature2013, Britnell_Science2013, Larentis_Nano2014} In regard to the practical applications, researchers have constructed electronic logic gates, transistors, memory devices, optical switches and biosensors using graphene/MoS$_2$ heterostructures. \cite{RoyNature2013,SimoneACSNano2013, SupNature2013, Britnell_Science2013, Larentis_Nano2014, Byungjin2015}

In this chapter, we review the structural, electronic and topological features of graphene/MoS$_2$ heterostructures. This chapter can be divided into two main parts: (i) Survey of results from the first-principles calculations, and (ii) Insights from the model Hamiltonian analysis and topological phase transitions. In the first part, we describe details regarding the crystal structure, interlayer effects, electronic bandstructure, nature of spin states and atomic orbitals near Fermi-level, strain effects on the electronic bandstructure, and charge-transfer phenomena. In the second part, we investigate the proximity effects and generic features of graphene/MoS$_2$ heterostructures using a tight binding formalism to obtain parameters for the symmetry-allowed low-energy effective Hamiltonian. Effects of the gate voltage on the dynamics of the bandstructure are discussed. Calculations of Berry curvature and Chern number confirm the occurrence of topological phase transitions at a critical gate voltage. The details of Density Functional Theory (DFT) calculations are given in the Appendix.



\section{Results from the DFT calculations}

\subsection{Insights into the graphene/MoS$_2$ heterostructure}

\begin{figure}[hb!]
 \centering
 \includegraphics[width=11.5cm, keepaspectratio=true]{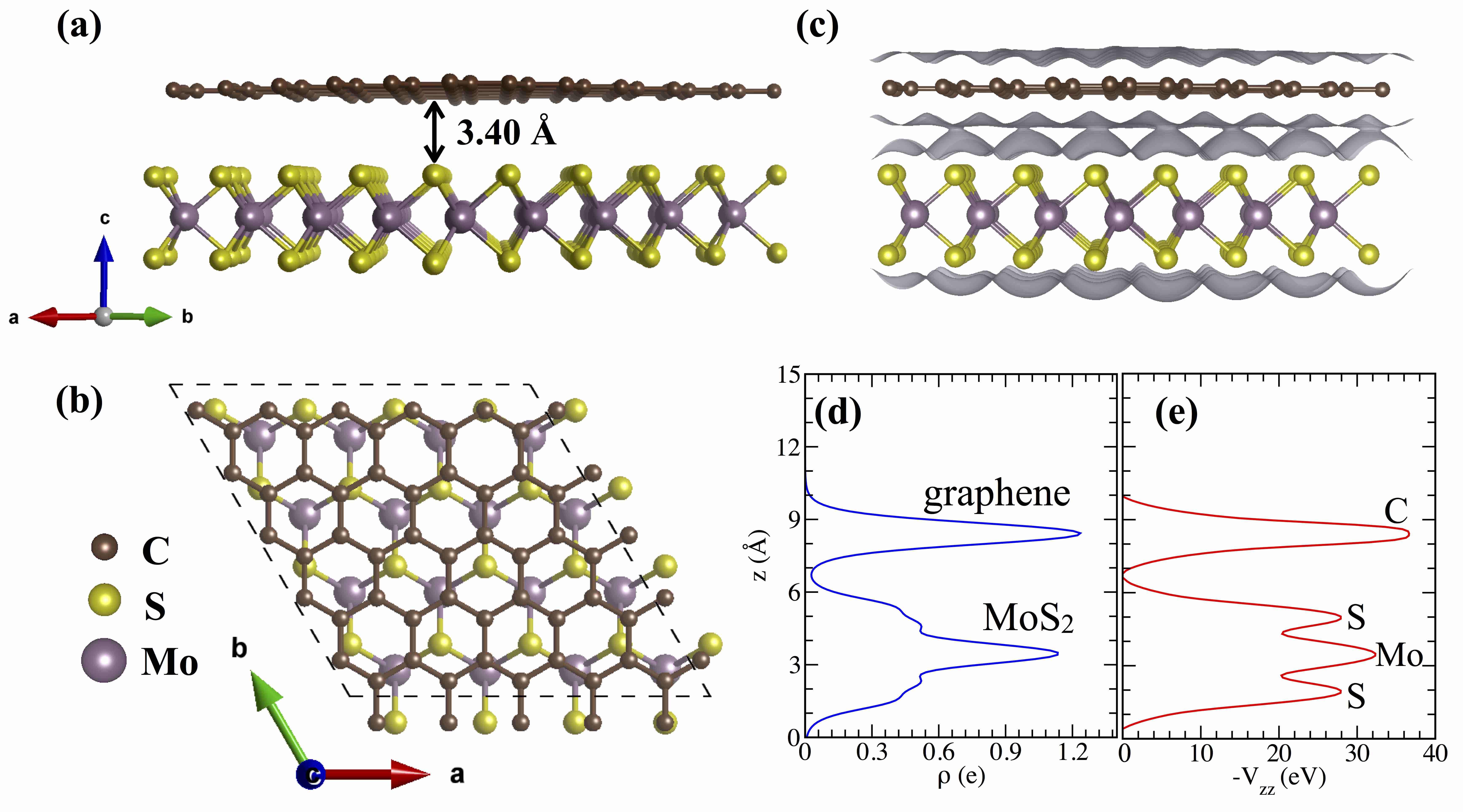}
 \caption{(Color online) Figures (a-b) show the crystal structure of 5:4 graphene/MoS$_2$ bilayer heterostructure from two different perspectives. (c) Isodensity charge surfaces (grey color) at isosurface level of $n$=0.007 for 5:4 graphene/MoS$_2$ bilayer heterostructure. The planar average of (d) charge density ($\rho$), and (e) electrostatic potential (V$_{zz}$) along the vertical $z$ direction. Notice the negative sign of V$_{zz}$ in figure (e).}
 \label{fig:crystal}
 \end{figure}

The optimized crystal structure of graphene/MoS$_2$ bilayer heterostructure is given in Fig. 1(a-b). Large lattice mismatch between graphene and MoS$_2$ monolayers makes the $ab$-$initio$ modeling of graphene/MoS$_2$ heterostructure computationally demanding. In order to minimize the lattice mismatch, one can vertically stack two commensurate supercells of graphene and monolayer MoS$_2$. Two most commonly used graphene/MoS$_2$ heterostructures are: (i) (4 $\times$ 4)/(3 $\times$ 3) (hereafter 4:3), and (ii) (5 $\times$ 5)/(4 $\times$ 4) (hereafter 5:4), where the latter has relatively smaller lattice mismatch but larger number of atoms/cell. In graphene/MoS$_2$ heterostructures, graphene and MoS$_2$ monolayers weakly interact through long-range vdW interactions. The experimentally reported interlayer distance between graphene and MoS$_2$ nanosheets is 3.40 \AA. \cite{PierucciNano2016} However, numerous $first$-$principles$ studies inconsistently predicted interlayer gap values ranging from 3.1 \AA~to 4.3 \AA. \cite{Abbas_APL2014, GmitraPRB2015, ShaoJPCC2015, NamLe2016, Hu2016, Jin2015, SachsAPL2013} This is mainly because of the inadequate evaluation of weak non-local vdW interactions within the DFT framework. Although, various DFT-vdW methods \cite{Grimme2006,LeePRB2010, Klime2011} have been employed and found to be inadequate in describing this system, it has been reported that the Tkatchenko-Scheffler (TS) method \cite{TS_PRL2009} for vdW corrections efficiently evaluates the long-range vdW interactions in this system, and accurately predicts the interlayer spacing (3.40 \AA) between graphene and MoS$_2$ nanosheets, \cite{Singh_gMoS2_2018} which is in remarkable agreement with the experimental data. The main reason behind the success of the TS method is the fact that it accounts for the non-local charge density fluctuations near the interface, whereas most of the other DFT-vdW methods are insensitive to the chemical environment. Therefore, it is expected that compared to other DFT-vdW methods, the TS-method might perform better in evaluating the weak vdW interaction between a metallic and an insulating material interface, where fluctuations in charge density are very large. \cite{Singh_gMoS2_2018}

The optimized lattice parameters of the 5:4 bilayer with minimal lattice mismatch are $a$ = $b$ = 12.443 \AA. \cite{Singh_gMoS2_2018} The Mo-S and C-C bond lengths are 2.38 and 1.44 \AA, respectively. In this case, the MoS$_2$ monolayer is being compressed by 0.3\%, whereas the graphene monolayer is being stretched by 1.16\% from the pristine case. The vertical distance between S-S atomic planes, {\it i.e.} the absolute thickness of the MoS$_2$ monolayer is 3.13 \AA. Fig. 1(c) shows the charge density isosurface near the interface. One can notice a small charge overlap between two constituent monolayers. This charge overlap is originating due to the weak vdW effects, and it could cause enhancement in the direct bandgap at Dirac point, as predicted by McCann. \cite{McCannPRB2006}. Variation in the planar average of charge density ($\rho$) and planar average of total local potential (V$_{zz}$) along the vertical $z$ direction is shown in Fig. 1(d-e). Here, V$_{zz}$ only includes the electrostatic part of potential without inclusion of the exchange-correlation term. Notably, there exists a potential difference between graphene and MoS$_2$ monolayers indicating presence of a non-zero dipole moment pointing towards the graphene layer. The amplitude of this dipole moment is $\sim$0.62 Debye in graphene/MoX$_2$ and $\sim$0.66 Debye in graphene/WX$_2$ heterostructures (X = S, Se). \cite{gmitra2017proximity}

\subsection{Electronic bandstructure: orbital and spin configurations}

\begin{figure}[hb!]
 \centering
 \includegraphics[width=11.5cm, keepaspectratio=true]{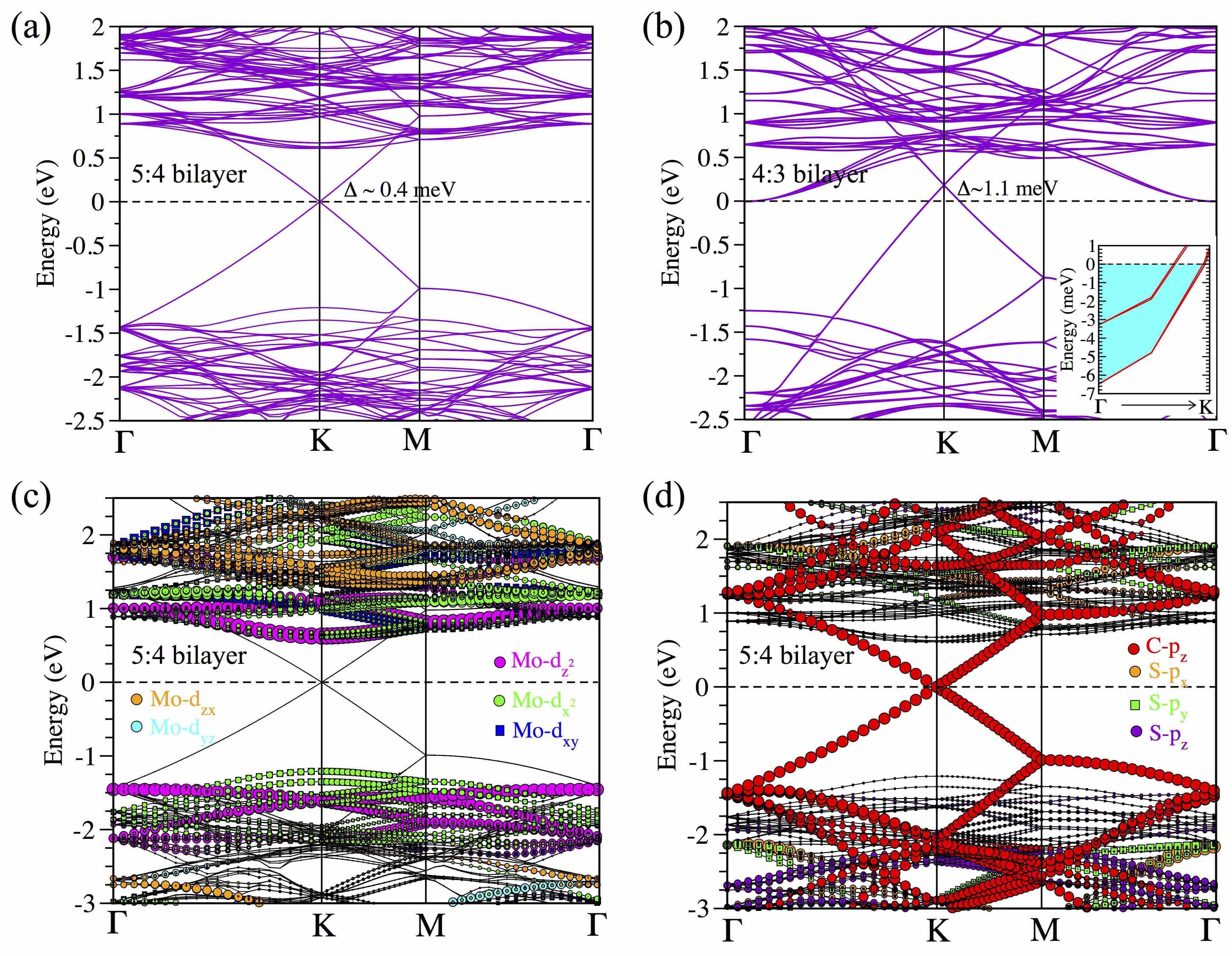}
 \caption{(Color online) The electronic bandstructures of (a) 5:4, and (b) 4:3 bilayer heterostructures calculated with vdW+SOC. Inset of Fig. (b) shows an enlarged view of the conduction bands near the $\Gamma$ point. Figures (c-d) represent the projection of atomic orbitals on the electronic bands of 5:4 bilayer. Horizontal dotted line at 0.0 eV energy marks the Fermi-level. }
 \label{fig:bands}
 \end{figure}

Fig.~\ref{fig:bands} shows the electronic bandstructure of two graphene/MoS$_2$ heterostructures (5:4 and 4:3) calculated with vdW+SOC along the high symmetry directions of the hexagonal Brillouin zone. The electronic features of graphene and MoS$_2$ monolayers are well preserved due to the weak vdW interaction between the monolayers. The linear dispersion of the Dirac cone lies within the bandgap of the MoS$_2$ monolayer in the 5:4 bilayer heterostructure. Contribution of various atomic orbitals to the electronic bands is shown in Fig~\ref{fig:bands}(c-d). Knowledge of the atomic orbitals near the Fermi-level is crucial for many theoretical and experimental investigations, such as: tight-binding calculations, determination of optical properties, charge carrier dynamics, photocatalysis, etc. Here, two notable features are: (i) the conduction and valence band of MoS$_2$ near the Fermi-level are mainly composed of Mo- $d_{z^2}$, $d_{xy}$ and $d_{x^{2} - y^{2}}$ orbitals, and (ii) the Dirac cone is formed by the $\pi$ bonded C-$p_z$ orbitals situated at A and B sublattices of graphene. The lowest conduction band near the Dirac point arises from the $p_z$ orbitals at the A-site, while the highest valence band arises from the $p_z$ orbitals at the B-site. All other states contribute to bands far from the Fermi-level as shown in Fig.~\ref{fig:bands}(c-d). \cite{Singh_gMoS2_2018}

The weak vdW interaction between graphene and MoS$_2$ monolayers yields a small, yet significant, bandgap at the Dirac point. The bandgap in 5:4 bilayer is $\sim$0.4 meV which increases almost by three times in 4:3 bilayer heterostructure due to the relatively larger lattice mismatch present in the 4:3 bilayer. Another interesting feature we observe in 4:3 bilayer heterostructure is the shift of the optical (direct) bandgap of MoS$_2$ monolayer from K to the $\Gamma$ point of Brillouin zone. In a 5:4 bilayer heterostructure, the MoS$_2$ monolayer preserves its direct bandgap semiconducting nature at the K-point with a direct bandgap of $\sim$1.8 eV, which is in excellent agreement with the reported values in the literature. \cite{Ramasubramaniam2012, Espejo2013, UdoPRB2015, Cecil2017} However in a 4:3 bilayer, the lowest conduction band shifts lower in energy at the $\Gamma$-point, whereas the highest valence band (at $\Gamma$-point) shifts higher in energy than the valence band maximum at the K-point. These two bands have Mo-$d_{z^2}$ character at $\Gamma$-point. Consequently, the direct energy gap of MoS$_2$ monolayer decreases in magnitude and shifts from the K-point to the $\Gamma$ point of Brillouin zone. Since the 5:4 graphene/MoS$_2$ bilayer heterostructure maintains the direct gap nature of MoS$_2$ monolayer at the K-point, it can be concluded that the aforementioned transition in 4:3 bilayer is primarily triggered by the strain effects arising due to the large lattice mismatch. \cite{Singh_gMoS2_2018}

Signatures of charge-transfer between the graphene and MoS$_2$ layers can be observed in Fig.~\ref{fig:bands}(b). The Dirac point in 4:3 bilayer is shifted above the Fermi-level and resides above the lowest conduction band with MoS$_2$ character. This indicates transfer of electrons from graphene to MoS$_2$ monolayer. This charge-transfer process can be harnessed by means of bi-axial strain or gate voltage, and is of central interest for technological applications. \cite{WeiHan2014, RoyNature2013} The net shift of Dirac point above the Fermi-level is $\sim$ 0.18 eV. Since the Dirac point has shifted above the Fermi level, the bottom of the conduction band of MoS$_2$ is expected to dip below the Fermi-level to catch the electrons transferred from graphene. In fact, a careful investigation of the lowest conduction band of MoS$_2$ near the Fermi-level shows that the Fermi-level is almost 6.5 meV above the bottom of the conduction band at the $\Gamma$-point, thus  suggesting the presence of an electron pocket at the $\Gamma$-point [see the inset of Fig.~\ref{fig:bands}(b)].

No such charge-transfer has been observed in 5:4 bilayer heterostructure which has minimal strain. This finding is consistent with the experimental observations of Diaz et al. \cite{DiazNano2015} In 2015, Diaz et al. performed angle-resolved photoemission spectroscopic (ARPES) measurements to probe the electronic structure of graphene/MoS$_2$ heterostructure. They observed that the Dirac cone of graphene remains intact and no significant charge-transfer occurs between the graphene and MoS$_2$ layers. However, bandgaps are reported away from the Dirac point due to the proximity of MoS$_2$. \cite{DiazNano2015}

\begin{figure}[hb!]
 \centering
 \includegraphics[width=11.5cm, keepaspectratio=true]{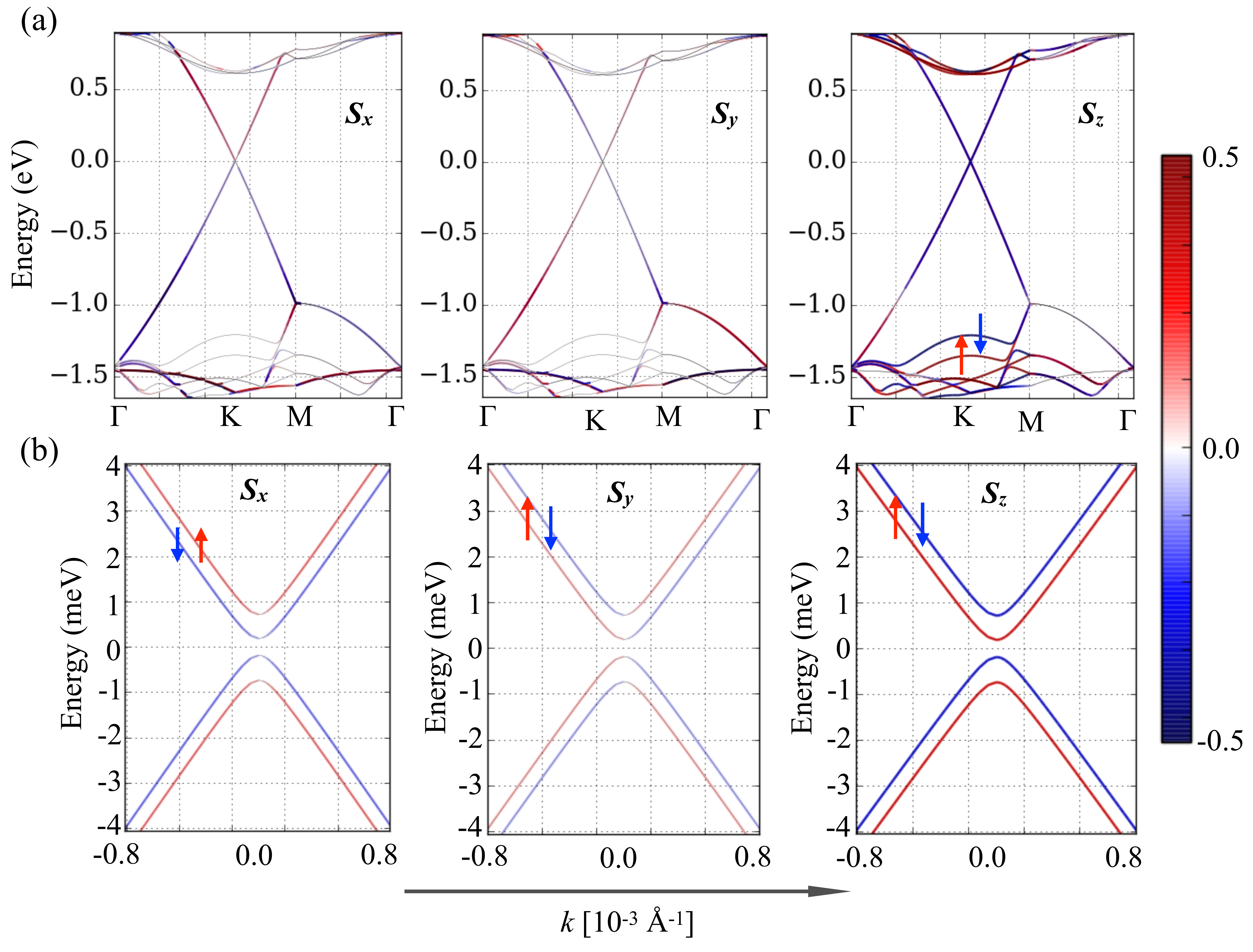}
 \caption{(Color online) Projection of $S_x$, $S_y$ and $S_z$ components of spin on the electronic bandstructure of the 5:4 bilayer heterostructure. Figures in the top panels (a) show various spin-contributions on the selected bands near the Fermi-level. Figures in the bottom panels (b) show the enlarged view of spin-splitting in bands near the Dirac point. The $k$-path in lower panels is centered at the hexagonal Brillouin zone K-point. Red (Blue) color depicts spin up (down) states.}
 \label{fig:spin}
 \end{figure}

After discussing the nature of orbitals and energy bandgap, we focus our attention on the spin related features of the electronic states in the graphene/MoS$_2$ bilayer. Figure~\ref{fig:spin} shows the projection of $S_x$, $S_y$ and $S_z$ components of spin on the electronic bandstructure of the 5:4 bilayer. Similar spin features are present for the 4:3 bilayer. The spin quantization axis was chosen along the (001) direction. As one can notice in Fig.~\ref{fig:spin}, the $S_z$ component of spin plays the dominant role in governing the spin features of bands near the Fermi-level, while the contribution of $S_x$ and $S_y$ projections is negligible. In the top panels of Fig.~\ref{fig:spin}, we plot the spin-projection on selected graphene and MoS$_2$ bands near Fermi-level, whereas the bottom panels show an enlarged view close to the neutrality point. In Fig.~\ref{fig:spin}(a), one can observe that Mo-$d$ top valence bands spin-split near the K-point due to the broken inversion symmetry (marked by red and blue arrows). The spin-splitting ($\Delta_{VB}$) is $\sim$0.2 eV at the K-point, which is not significantly affected by the nearby graphene layer. Notice this is much smaller than that reported for WX$_2$ monolayers (X = S, Se, Te). The value of $\Delta_{VB}$ for WS$_2$, WSe$_2$, and WTe$_2$ is 0.43 eV, 0.47 eV, and 0.48 eV, respectively. \cite{JKang2013, Amin2014} This is as expected from the difference in the atomic numbers of S, Se, and Te.

An enlarged view of bands near the Fermi energy reveals that bands acquire a parabolic shape near the Dirac point due to proximity effects. A Rashba-type spin-splitting is expected in this system because of the broken inversion symmetry and strong SOC effects arising from the MoS$_2$ layer. Moreover, due to the intrinsic SOC of graphene, a spin-gap opens at the Dirac point and bands anticross each other yielding the resulting band dispersion shown in Fig.~\ref{fig:spin}(b). \cite{Singh_gMoS2_2018} Staggered potential effects further enhance the bandgap opening. By harnessing the aforementioned competitive terms, one can realize distinct topological phases in this bilayer system. \cite{Abdulrhman2016} A controlled phase transition between the distinct topological phases can be achieved either by tuning strength of SOC from the TMDC layer or by applying a relative gate voltage between the layers. \cite{Abdulrhman2016} We discuss this issue in more detail later using a model Hamiltonian.

\subsection{Strain effects and charge transfer}

As we mentioned above while discussing the electronic bandstructure of 5:4 and 4:3 bilayer heterostructures, the shifting of the Dirac point above the lowest conduction band of MoS$_2$ indicates the occurrence of a charge-transfer from graphene to the MoS$_2$ monolayer. We also argued that this charge-transfer is mainly triggered by strain. The effect of strain on the electronic properties of graphene \cite{ZhenNi2008, FerralisPRL2008, MohrPRB2009, PereiraPRB2009, ChoiPRB2010, guinea2010energy, C5NR07755A, SharmaPRB2017} and MoS$_2$ \cite{HuiPan2012, C2CP42181J, GhorbaniPRB2013, Cappelluti2013, Conley2013, HongliangPRB2013, Keliang2013,Scalise2012_NanoR, Castellanos2013, Swastibrata2014, SCALISE2014416,Feierabend2017} has been well evaluated in the literature from both theoretical and experimental studies. These studies conclude that the electronic properties of both graphene and MoS$_2$ monolayer can be considerably harnessed by strain engineering and novel features can be realized in these monolayer systems. At moderate strains, graphene maintains its semimetallic feature. No significant changes in the electronic bandstructure of graphene have been observed for strains up to $\sim$15\%. However, depending upon the magnitude and direction of applied strain, Dirac cone can be shifted away from the K point. Choi et al. \cite{ChoiPRB2010} predicted that no sizable energy gap opens in the uniaxially strained graphene under uniaxial strain less than 26\% along any arbitrary direction. They further suggested that the low-energy dispersion of bands in moderately uniaxially-strained graphene can be modeled using the generalized Weyl's equation. \cite{ChoiPRB2010} As the uniaxial strain increases, the Fermi-velocity of Dirac cone varies (increases or decreases) depending upon the direction of the wave vector. \cite{ChoiPRB2010} Interestingly, Guinea et al. \cite{guinea2010energy} have reported that a designed strain aligned along three main crystallographic directions could induce strong gauge fields which effectively act as a uniform pseudomagnetic field. 

On the other hand, at a critical value of strain, the valence band maxima of MoS$_2$ at $\Gamma$ increases in energy, shifting towards the Fermi-level, and supersedes the valence band maxima of MoS$_2$ at K, thus resulting in a direct to indirect bandgap transition in the strained monolayer. A number of theoretical as well as experimental studies have concluded that this bandgap transition occurs in MoS$_2$ at 0.5--1.0\% compressive or tensile strain. \cite{HuiPan2012, C2CP42181J, GhorbaniPRB2013, Cappelluti2013, Conley2013, HongliangPRB2013, Keliang2013,Scalise2012_NanoR, Castellanos2013, Swastibrata2014, SCALISE2014416,Feierabend2017} Considering many-body and SOC effects, Wang et al. \cite{ANDP201400098} predicted that the direct to indirect gap transition in MoS$_2$ monolayer should occur at 2.7\% strain. \cite{SCALISE2014416} Under a tensile strain, the thickness of the MoS$_2$ monolayer ({\it i.e.} separation between S-S planes) decreases owing to its positive Poisson's ratio, \cite{YUE20121166} which results in enhanced hybridization of S-$p_z$ orbitals that contribute to the valence band maxima at $\Gamma$. However, Mo-$d_{z^2}$ orbitals mostly remain unaltered under the biaxial strain conditions, while Mo-$d_{xy}$ and Mo-$d{_{{x^2}-{y^2}}}$ states suffer energy shifts when strain is imposed. Such strain-induced direct to indirect bandgap transition manifests as decreasing photoluminescence intensity of MoS$_2$ monolayer and it can be clearly traced in experiments. \cite{Conley2013} The energy bandgap of MoS$_2$ decreases upon application of strain. Moreover, the effective mass of electrons and holes at K and $\Gamma$ points decreases with increasing strain. \cite{SCALISE2014416, ANDP201400098} The rate of reduction for hole effective mass at $\Gamma$ is much higher compared to the reduction of electron effective mass at K. For instance, the effective mass for holes is reduced by more than 60\% at $\Gamma$, while the effective mass of electrons at K drops by 25\% for a tensile strain of 5\%. \cite{SCALISE2014416} Interestingly, a semiconductor to metal transition is predicted in MoS$_2$ monolayer at a tensile strain of $\sim$10\% and at a compressive strain of $\sim$15\%. \cite{SCALISE2014416}

Notably, the direct to indirect bandgap transition in MoS$_2$ can also be achieved by vertically stacking two or more monolayers. With increasing number of layers, the interaction between the Mo-$d_{z^2}$ orbitals of different S-Mo-S nanosheets increases which leads to an upshift of the energy bands. Consequently, the valence band maximum at $\Gamma$ and conduction band minimum at K shift towards higher energy values, whereas other states do not change much being mainly composed of $d$ orbitals lying in $x-y$ plane. For this reason, multilayer MoS$_2$ exhibits an indirect bandgap between the valence band maximum at $\Gamma$ and the conduction band minimum along the $\Gamma - K$ path. \cite{SCALISE2014416}

\begin{figure}[hb!]
 \centering
 \includegraphics[width=11.5cm, keepaspectratio=true]{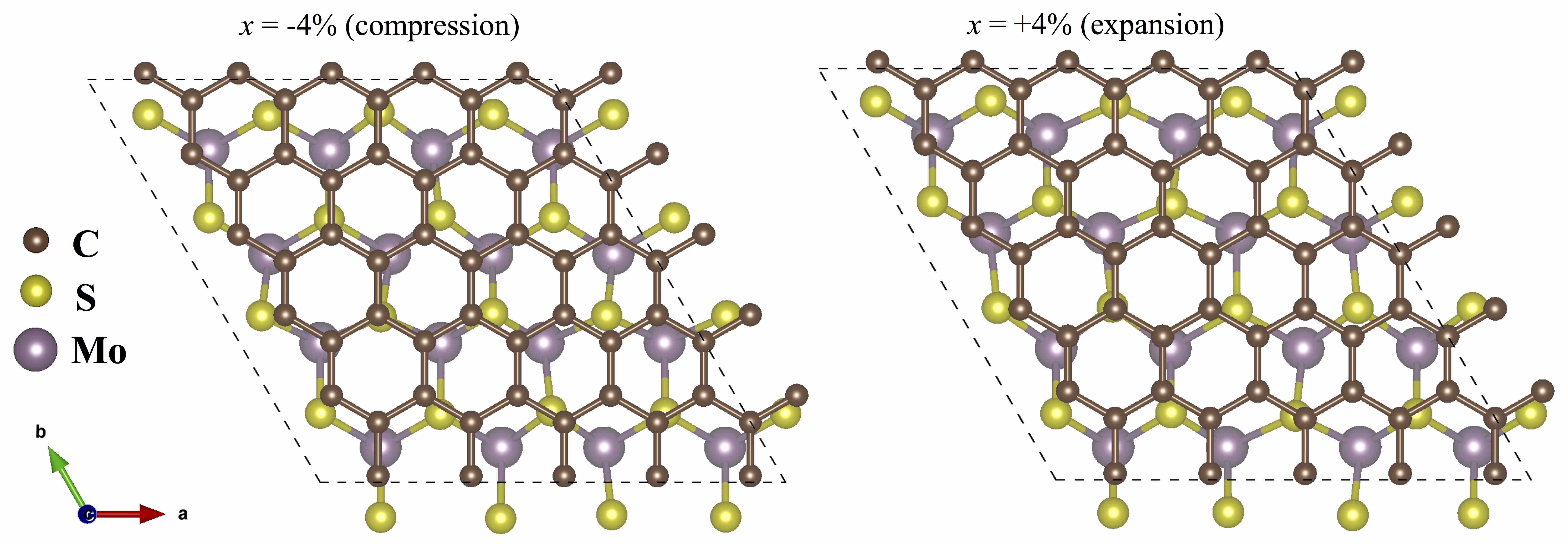}
 \caption{(Color online) Figures show the crystal structure of biaxially strained 5:4 graphene/MoS$_2$ bilayer from the top view. Left panel represents the case when Mo atoms are compressed by $4\%$ ({\it i.e} $x = -4\%$) while right panel represents the case when Mo atoms are expanded by $+4\%$ ({\it i.e} $x = +4\%$).}
 \label{fig:crystal_strain}
 \end{figure}

In the simplest approximation, it can be assumed that Mo atoms primarily suffer the interfacial strains caused by the substrate, whereas S atoms relax according to the modified location of the strained Mo atoms. Here, we perform a computational exercise to understand the effect of biaxial strain on Mo atoms on the electronic structure of graphene/MoS$_2$ heterostructure. We apply biaxial strain on Mo atoms in the well optimized 5:4 graphene/MoS$_2$ bilayer heterostructure and fully relax the S atoms in the strained cell. Biaxial strain ($x$) ranging from $-4\%$ (compressive strain) to $+4\%$ (expansion or tensile strain) was employed on Mo atoms. This computational exercise roughly models the local substrate induced strain effects on the Mo atoms which disrupt the ordering of Mo atoms in lattice yielding formation of domains or grain boundaries at finite intervals. In our case, grain boundaries would be formed at the edge of the unit cell of dimensions: $a=b$ = 12.44 \AA, where two Mo atoms from adjacent periodic cells would either come close to each other or move away depending upon the tensile or compressive strains employed on the Mo atoms, respectively. Fig.~\ref{fig:crystal_strain} shows the crystal structure of strained 5:4 graphene/MoS$_2$ bilayer heterostructure for two extreme cases of employed biaxial strain ($x$) on Mo atoms. Positive/Negative values indicate the tensile/compressive strain. We observe a small increase in the absolute thickness of MoS$_2$ monolayer with increasing compressive strain which is as expected due to the positive Poisson's ratio of MoS$_2$ monolayer.  \cite{YUE20121166} Because of the weak vdW interaction between graphene and MoS$_2$ nanosheets, we notice a negligible change in the interlayer separation with varying $x$, which is consistent with changing MoS$_2$ thickness. The maximum change in interlayer distance is $\pm$0.02 \AA~at the extreme values of imposed strains on Mo atoms.

\begin{figure}[hb!]
 \centering
 \includegraphics[width=11.5cm, keepaspectratio=true]{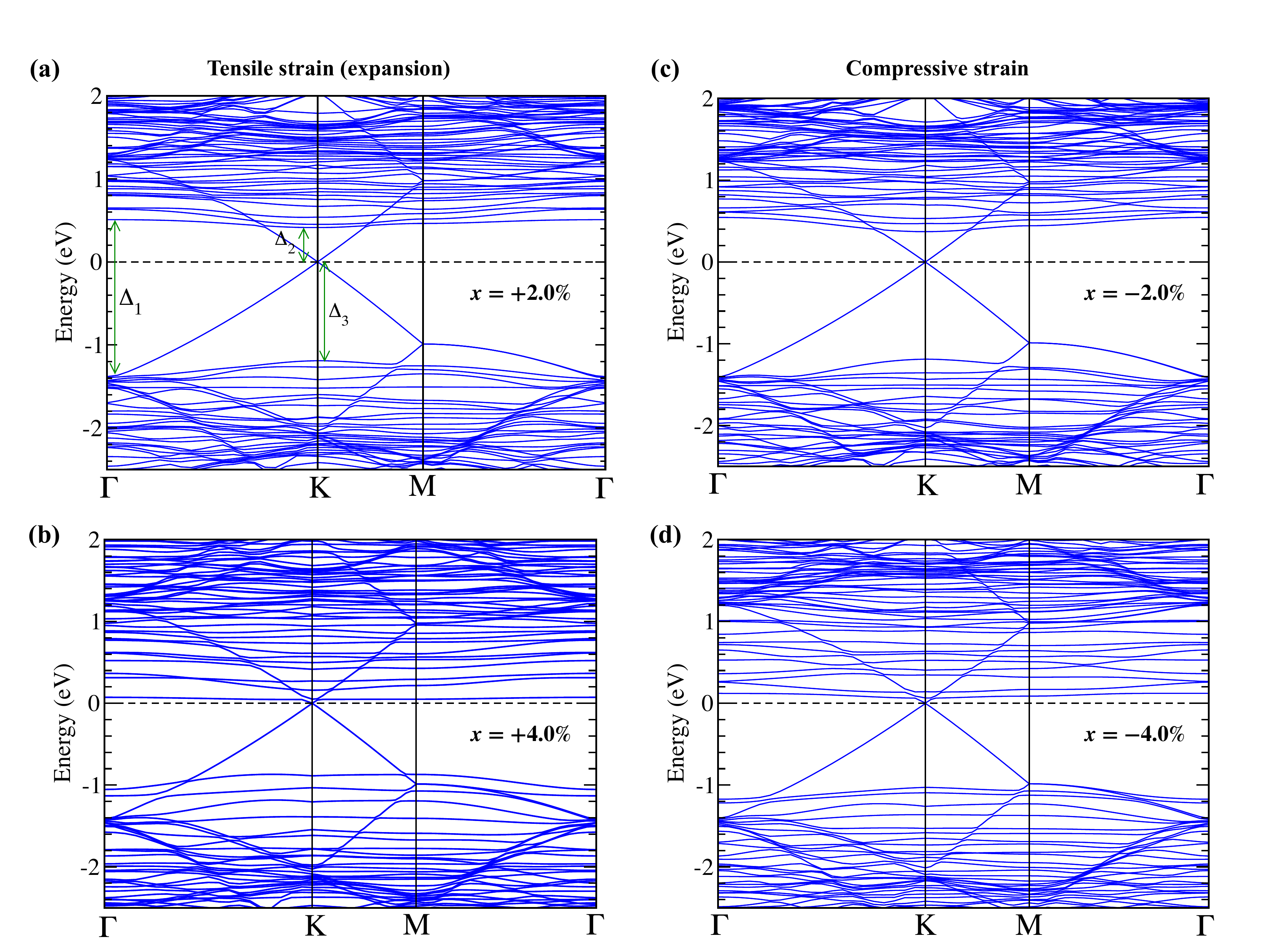}
 \caption{(Color online) Figures show the electronic bandstructure of strained Mo atoms in 5:4 graphene/MoS$_2$ bilayer heterostructure calculated without SOC. Fig. (a) and (b) represent bands for $2.0\%$ and $4.0\%$ tensile strains, whereas Fig. (c) and (d) represent bands for $2.0\%$ and $4.0\%$ compressive strains, respectively. }
 \label{fig:band_strain}
 \end{figure}

Fig.~\ref{fig:band_strain} shows the electronic bandstructure of 5:4 graphene/MoS$_2$ bilayer having strained Mo atoms. Both compressive and tensile strains yield similar features in the electronic bands. With increasing strain on Mo atoms, both valence and conduction Mo-$d$ bands shift towards the Fermi-level decreasing the net bandgap of the MoS$_2$ monolayer. However, MoS$_2$ maintains the direct bandgap nature in the studied range of strain. This finding is important since it suggests that graphene/MoS$_2$ heterostructure mounted on a suitable substrate that imposes small interfacial strain on Mo atoms can be considerably tuned by controlling the substrate-imposed strain on Mo atoms. This effect can be present in photoluminescence experiments. \cite{MoS2onSiO2, MoS2Galli2} One can also notice that the effective mass of charge carriers in MoS$_2$ monolayer increases with increasing strain on Mo atoms.

\begin{figure}[hb!]
 \centering
 \includegraphics[width=11.5cm, keepaspectratio=true]{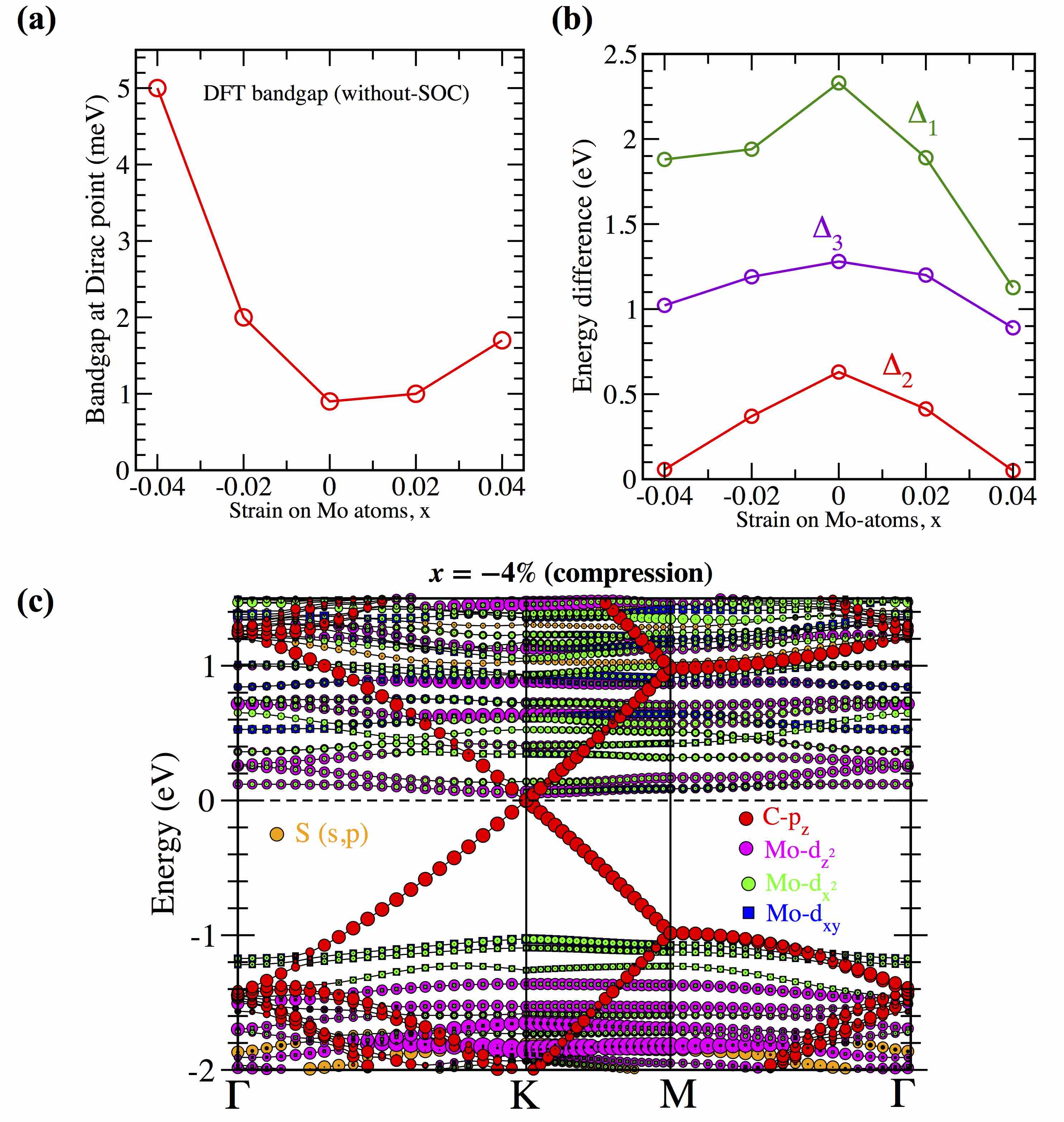}
 \caption{(Color online) Figure (a) and (b) represent change in the direct bandgap at Dirac point, and quantities $\Delta_1$, $\Delta_2$, and $\Delta_3$ as a function of the strain on Mo atoms -- $x$. See Fig.~\ref{fig:band_strain}(a) for definition of $\Delta_1$, $\Delta_2$, and $\Delta_3$. (c) Projection of the selected atomic orbitals on the electronic bands of 5:4 graphene/MoS$_2$ bilayer heterostructure having 4\% compressively strained Mo-atoms. This bandstructure was calculated without inclusion of SOC.} 
 \label{fig:gap_strain}
 \end{figure}

In order to further understand the effect of strain on the direct bandgap at Dirac point, location of band edges of MoS$_2$ monolayer, and change in the orbital features near the Fermi-level, we plot the aforementioned quantities as a function of $x$ in Fig.~\ref{fig:gap_strain}. Projection of various atomic orbitals on the electronic bands for $x = -4\%$ case reveals the nature of orbitals near the Fermi-level is preserved in the studied range of imposed strain on Mo atoms. The direct bandgap at Dirac point increases substantially with increasing strain on Mo atoms [see Fig.~\ref{fig:gap_strain}(a)]. This can be attributed to the enhanced hybridization between $d_z$ and $p_z$ orbitals. Fig.~\ref{fig:gap_strain}(b) shows variation in $\Delta_1$, $\Delta_2$, and $\Delta_3$ versus $x$. Here, $\Delta_1$ represents the energy difference between the lowest conduction and highest valence bands at $\Gamma$, $\Delta_2$ refers to the energy difference between the lowest conduction band of Mo-$d$ states and Dirac point, and $\Delta_3$ represents that between the Dirac point and the highest valence band of Mo-$d$ states [see Fig.~\ref{fig:band_strain}(a) for illustration]. Our analysis shows $\Delta_1$, $\Delta_2$, and $\Delta_3$ decreasing with increasing strain on Mo atoms. With increasing $x$, the Dirac point comes closer to the conduction bands of MoS$_2$, and at $x = \pm4\%$ the lowest conduction band of Mo-$d$ states almost touches the Dirac point. Therefore, beyond $x = \pm4\%$ strain, a charge-transfer may occur from graphene to MoS$_2$ monolayer. 

From the above discussion, it can be concluded that by tuning the substrate-induced strain on Mo-atoms, one can harness the optical properties of graphene/MoS$_2$ bilayer heterostructure and further control the charge-transfer process between the two monolayers. From an experimental perspective, this can be achieved by choosing a suitable piezoelectric or flexoelectric substrate.

\section{Model Hamiltonian and topological phase transitions}

\subsection{Basic theoretical model}
\begin{figure}[hb!]
	\centering
		\includegraphics[width=1.0\linewidth]{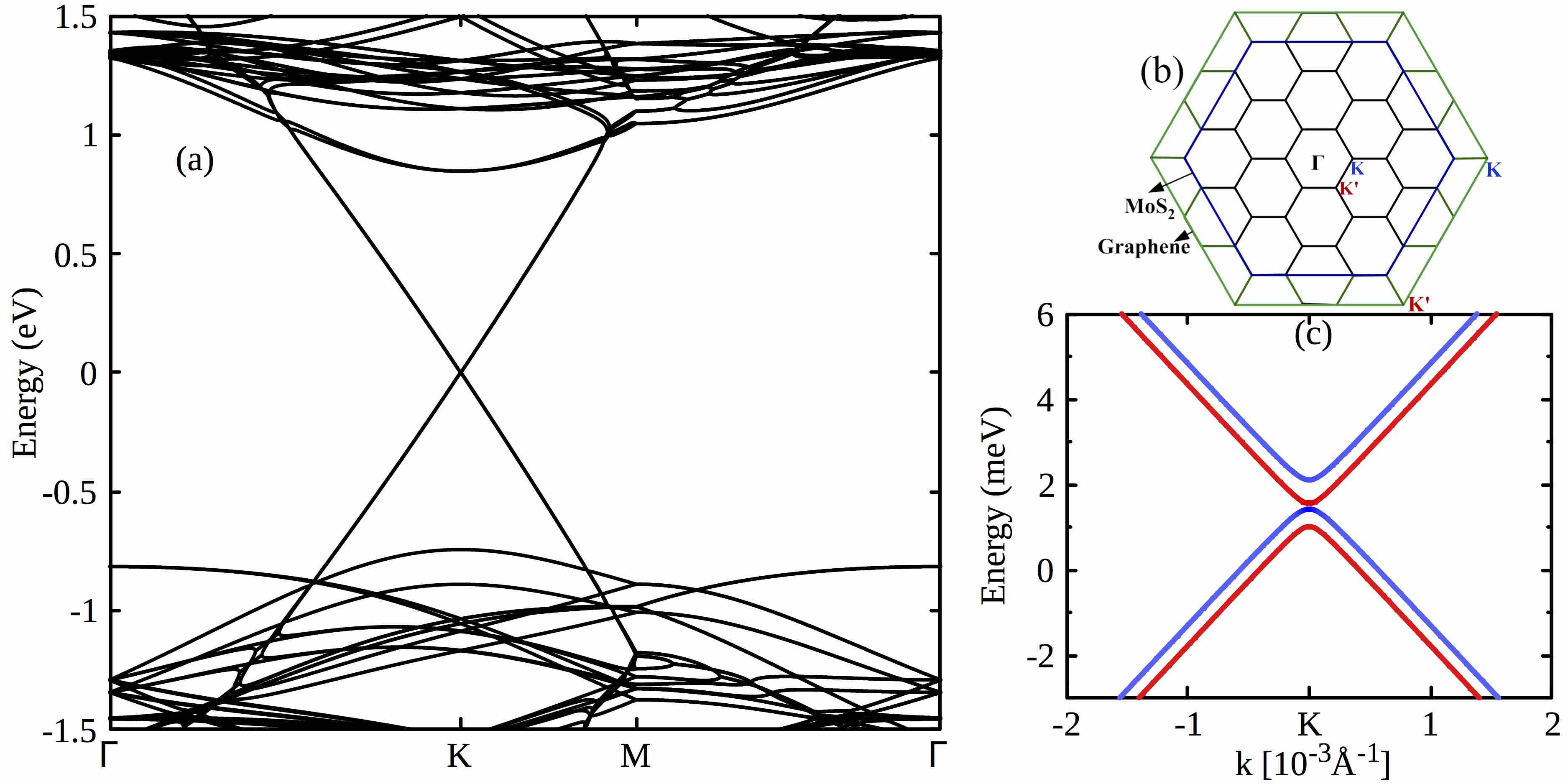}
	\caption{(Color online) Graphene/MoS$_2$ heterostructure in tight-binding description. (a) Band dispersion of graphene/MoS$_2$ along high symmetry lines
		$\Gamma$-K-M-$\Gamma$. (b) Brillouin zones of the reciprocal lattices. (c) Zooming near K valley shows that graphene bands are gapped and spin polarized due to the
		proximity of MoS$_2$. Blue (red) bands represent spin down (up) states.
		A graphene and MoS$_2$ monolayer first Brillouin zones (BZ) are shown as green and blue hexagon, respectively. Their relative K and K' valleys are also shown. The supercell BZ has a smaller reciprocal lattice size which upon folding, 
		maps corner valleys from both layers onto the same point \cite{Abdulrhman2016}. Compare this figure with Fig.\ \ref{fig:bands}, showing similar features, although here the Fermi level is symmetric in the TMD gap. }
	\label{Fig1}
\end{figure}

 In this section, we study the heterostructure using a tight-binding theoretical framework. First, a linear transformation that connects the primitive lattice vectors of graphene and MoS$_2$ is written as \cite{mori1},
\begin{equation}\label{vec.conn}
\left( 
\begin{matrix}
\textbf{a}_{G_1} \\
\textbf{a}_{G_2}
\end{matrix}
\right) 
={M}\cdot
\left( 
\begin{matrix}
\textbf{a}_{\rm {M}_1} \\
\textbf{a}_{\rm{M}_2}
\end{matrix}
\right) 
\text{ , }
\end{equation}
where $ \textbf{a}_{x_1}=a_x\left( \frac{\sqrt{3}}{2},\frac{1}{2}\right)$ and $\textbf{a}_{x_2}=a_x \left( \frac{\sqrt{3}}{2},\frac{-1}{2}\right)$ are the real primitive vectors (for $x$ = graphene and MoS$_2$) with ${M}= diag(\frac{4}{5},\frac{4}{5})$. 
It can be shown \cite{mori1} that the resulting moir\'{e} pattern has primitive lattice vectors  ($ \textbf{R}_{1} $ and $ \textbf{R}_{2} $) given by
\begin{equation}\label{R-vec}
\left( 
\begin{matrix}
\textbf{R}_{1} \\
\textbf{R}_{2}
\end{matrix}
\right) 
={[{1}-{M}]^{-1}{M}}\cdot
\left( 
\begin{matrix}
\textbf{a}_{\rm{M}_1} \\
\textbf{a}_{\rm{M}_2}
\end{matrix}
\right) 
\end{equation}
Because of the honeycomb structure, the Brillouin zone of the graphene/MoS$_2$ $4:5$ heterostructure has similar features to graphene, with two valleys $\textbf{K}=\frac{2\pi}{a_\alpha}\left( \frac{1}{\sqrt{3}},\frac{1}{3}\right)$ ,  $\textbf{K}' =\frac{2\pi}{a_\alpha}\left( \frac{1}{\sqrt{3}},\frac{-1}{3}\right)$, $a_\alpha=5 a_G = 4 a_{\rm{M}}$, where
the K and K$ ' $ valleys of graphene and MoS$_2$ are mapped onto the same positions of the first Brillouin zone of the supercell upon folding [see Fig.~\ref{Fig1}(c)].

We use a tight-binding formalism that couples up to next nearest neighbors $\langle\langle ij \rangle\rangle$ in MoS$_2$ with minimal three-orbital basis. In MoS$_2$, the basis can be represented at low-energy by three orbitals ($d_{z^2}$, $d_{xy}$ and $d_{x^2-y^2}$), as discussed above and in Ref. \cite{mos23}, so that
\begin{equation}\label{mo-h}
\begin{split}
\mathcal{H}_{{M}}=\sum_{i,\sigma,\nu}\epsilon_{\nu,\sigma} \alpha^{\dagger}_{i\nu\sigma}\alpha_{i\nu\sigma}  
+\sum_{\langle\langle ij\rangle\rangle, \nu\mu,\sigma} t_{i\nu,j\mu} \alpha_{i\nu\sigma}^{\dagger}\alpha_{j\mu\sigma}+h.c.,
\end{split}
\end{equation}
where $\alpha^{\dagger}_{j\nu\sigma}$ label the $\nu$-orbital at site $j$ of the Mo-lattice with spin $\sigma$. The first term considers the on-site energy of atom $j$ and orbital $\nu$. The second term describes hopping between Mo orbitals to nearest and next nearest neighbors. Strong MoS$_2$ spin orbit coupling is considered from atomic SOC contribution, (see Eq.\ 25 and Table IV in Ref.\ \cite{mos23}).

To model graphene, we adopt the usual single-orbital representation for the triangular lattice with two-atom basis that couples only nearest neighbors $\langle ij \rangle$ \cite{graphene},
\begin{equation}\label{G-h}
\begin{split}
\mathcal{H}_{{G}}=\sum_{i,\sigma}\epsilon_{i,\sigma} {c}^{\dagger}_{i\sigma}{c}_{i\sigma}
-t_{g}\sum_{\langle ij \rangle,\sigma} ({c}_{i\sigma}^{\dagger}{c}_{j\sigma}+h.c.),
\end{split}
\end{equation}%
where $\epsilon$ of the first term describes the on-site energy, and the second term considers hoppings to the nearest neighbors with coupling strength  $t_g$.

The presence of a substrate generates a perpendicular electric field to the graphene layer. This electric field causes a spin orbit coupling that can be described by a Rashba Hamiltonian of the form \cite{qshe}
\begin{equation}
\mathcal{H}_{R}= i t_R\sum_{\langle ij\rangle;\alpha ,\beta} \hat{\boldmath{e}}\cdot(\textbf{\textit{s}}_{\alpha\beta}\times \textbf{\textit{d}}^{\circ}_{ij}) c^{\dagger}_{i\alpha}c_{j\beta},
\end{equation}
where $\alpha,\beta$ describes spin up and spin down states, $\textbf{\textit{d}}^{\circ}_{ij}=  \frac{\textbf{\textit{d}}_{ij}}{\arrowvert\textbf{\textit{d}}_{ij}\arrowvert} $ is the unit vector that connects A atom of graphene to its nearest neighbor B atom. The Rashba spin orbit interaction is weak in graphene i.e., $t_R$ = 0.067 meV \cite{G-Rash}. This captures the mirror symmetry breaking effect. As a consequence, the spin is no longer a good quantum number and spin states interact with each other, opening anti-crossings at degeneracy points. 

We consider coupling only between neighbors across the layers between the graphene $ p_z $-orbital and MoS$_2$ $ d$-orbitals, which is described as
\begin{equation}\label{T-B-h}
\begin{split}
\mathcal{H}=\sum_{\langle ij\rangle,\nu\sigma} t^{\nu}_{i,j} {c}_{i\sigma}^{\dagger}{\alpha}_{j\nu\sigma}+h.c.
\end{split}
\end{equation}
where $t^{\nu}_{i,j}$ is represented by a tunneling amplitude
\begin{equation}\label{tunnel}
t^{\nu}_{i,j}=t_{\nu} \text{ exp}\left[ -\arrowvert \textbf{r}_{m,i}-\textbf{r}_{g,j}\arrowvert/\eta\right] ,
\end{equation}
where $\arrowvert \textbf{r}_{m,i}-\textbf{r}_{g,j}\arrowvert$ is the distance that connects atoms in both layers, normalized to a constant $\eta=5a_g$.  $t_{\nu}$ describes the effective coupling between $p_z$ and $d$-orbitals using a Slater-Koster approach \cite{slater}.  It takes the form \cite{Abdulrhman2016}
\begin{equation}\label{s-k}
\begin{split}
t^{z^2} &=\langle p_z|H|d_{z^2}\rangle \\
&=-\sqrt{3} n^{3}_{z} V_{pd\pi}-\frac{1}{2}n_{z}(n^{2}_{x}+n^{2}_{y}-2	n^{2}_{z})V_{pd\sigma} \\
t^{x^2-y^2} &=\langle p_z|H|d_{x^2-y^2}\rangle\\
&=\frac{\sqrt{3}}{2}(n_zn^2_x−n^2_y) V_{pd\sigma}-(n_zn^2_x−n^2_y)V_{pd\pi}\\
t^{xy}& =\langle p_z|H|d_{xy}\rangle\\
&=n_xn_yn_z(\sqrt{3}V_{pd\sigma}-2V_{pd\pi}),\\
\end{split}
\end{equation}
where $n_i$ are  directional cosines. 
The numerical values of the coupling constants are set to be in agreement with what is expected: the coupling $ t^{z^2} $ is larger than $t^{xy} $ and  $t^{x^2-y^2} $, due to a higher overlap. The numerical values used here, $V_{pd\pi}=-0.232$ eV and $V_{pd\sigma}=0.058$ eV, do not affect the main conclusions nor qualitative behavior, as we will discuss below. This Hamiltonian is capable of reproducing the low energy dispersions close to the K and K$ ' $ points with great accuracy. TMD parameters are adapted from Liu \textit{et al.} \cite{mos23}, while for graphene we take the on-site energy to be zero and hopping parameter $t_g$ = 3.03 eV \cite{graphene}.

\subsection{Dirac cone and gate voltage effects}
The full band structure of the heterostructure along high symmetry lines ($\Gamma$-M-K-$\Gamma$) is shown in Fig.\ \ref{Fig1}(a). A closer look near the Dirac points in Fig.\ \ref{Fig1}(b), shows that a gap in the bulk system appears, and that the spin degeneracies lifted, reflecting the broken inversion symmetry of the heterostructure. The gap size is (nearly quadratically) dependent on the interlayer tunneling parameters to chosen. The gap here is in the order of one meV. All the bandstructure features are the same as these seen in the first principals results shown in Fig.\ \ref{fig:bands}

We have discussed the charge transfer between layers in the previous section, where relative band alignment between graphene's Dirac point and low energy TMDs bands in this two layered system depends on strain and has been determined in calculations \cite{Band-Alignment1, Band-Alignment2,G-TMDs}. In the tight-binding model, the relative band alignment can be shifted to mimic applying an effective potential across layers \cite{Abdulrhman2016}. Hence, the relative position of graphene's Dirac points to the MoS$_2$ bands can be adjusted to study the effect of different orbital couplings, even if there is no charge transfer, using 
\begin{equation}\label{voltdrop}
\mathcal{H}_{\rm Gate}=V_{\rm Gate}\sum_{i\alpha} c^{\dagger}_{i\alpha}c_{i\alpha}.
\end{equation}

\begin{figure}[H]
	\includegraphics[width=11.5cm, keepaspectratio=true]{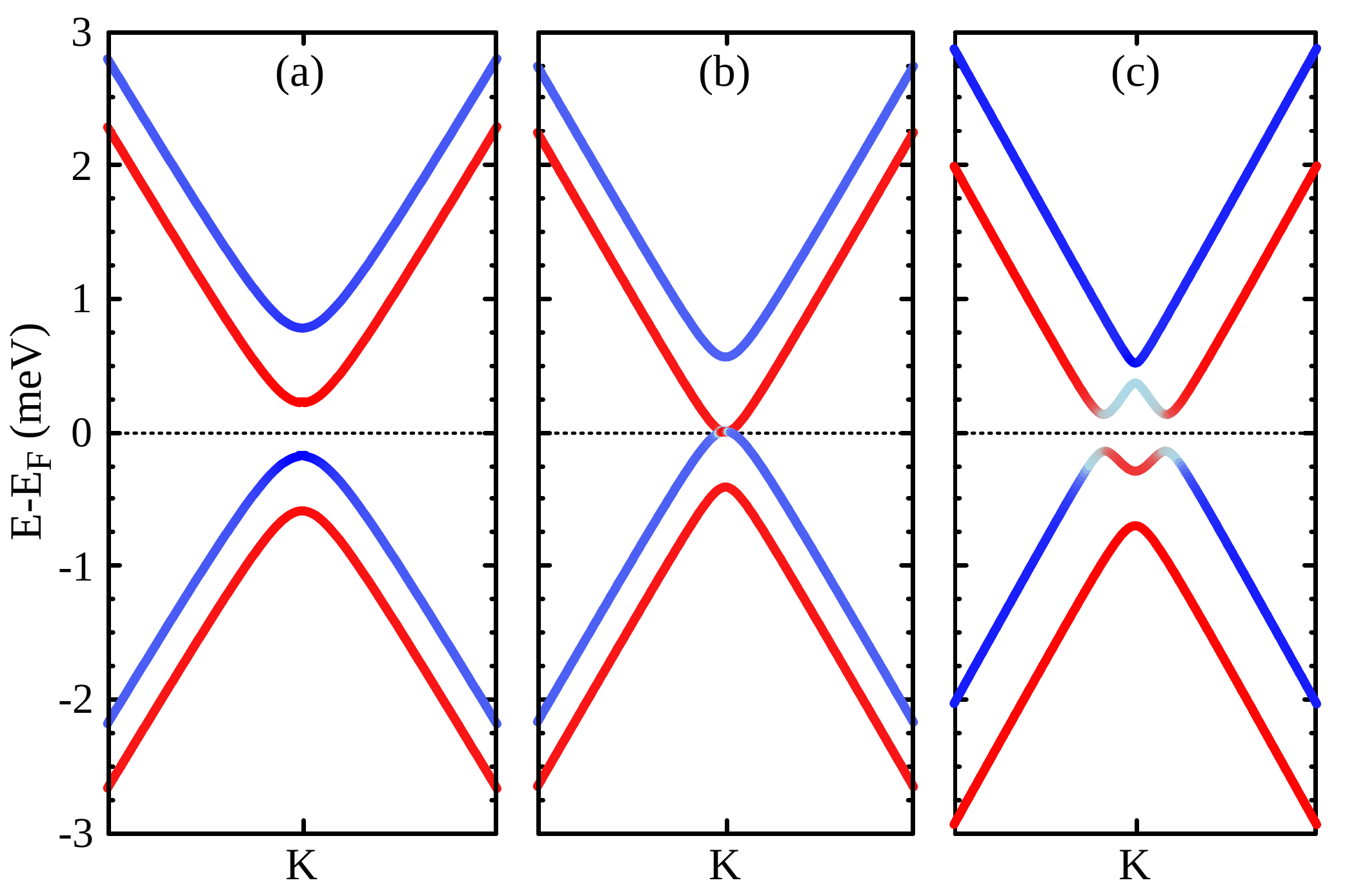}
	\caption{(Color online) Band structure of tight-binding model near the K valley for different gate voltages.  (a) Direct band regime with finite bulk gap where the graphene Dirac point is close to the conduction band of the TMD. (b) Semi-metallic phase with spin states split. (c) Inverted bands as the Dirac point of graphene is moved close to the valence bands of TMDs.  Red(blue) color describes spin up(down) states. States at K$'$ valley are spin reversed \cite{Abdulrhman2016}. }  
	\label{Fig2}
\end{figure}

As the MoS$_2$ low energy conduction and valence bands show different SOC and orbital characteristics. Tuning graphene's potential relative to the MoS$_2$ may then result-in different proximity effects onto the Dirac bands. Results of such tuning yields three distinct states as shown in Fig.~\ref{Fig2}. A large direct gap band is generated due to proximity to the conduction band of MoS$_2$ [see Fig.~\ref{Fig2}(a)]. The band gap is proportional to the gate voltage applied, increasing as we increase the gate voltage. In contrast, proximity of the Dirac point to valence bands of MoS$_2$ yields an inverted band gap, shown in Fig.~\ref{Fig2}(c). This inverted gap band is produced due to the interaction between bands caused by Rashba spin orbit coupling. The bands around the Fermi level have mixed spin states (spin up and down) in the case of inverted band gaps, while they show definite spin states in the direct gap band phase. These two topological phases are separated by a semi-metallic gapless state, shown in Fig.~\ref{Fig2}(b).

Notice that the spin state in K is reversed in the  K$ ' $ valley due to time reversal symmetry, whereas the staggered potential value is the same for both valleys. Further analysis to study the changes of topological characters accompanied with the change of these band phases is discussed below.

\subsection{Spin State}
\begin{figure}[hb!]
	\includegraphics[width=11.5cm, keepaspectratio=true]{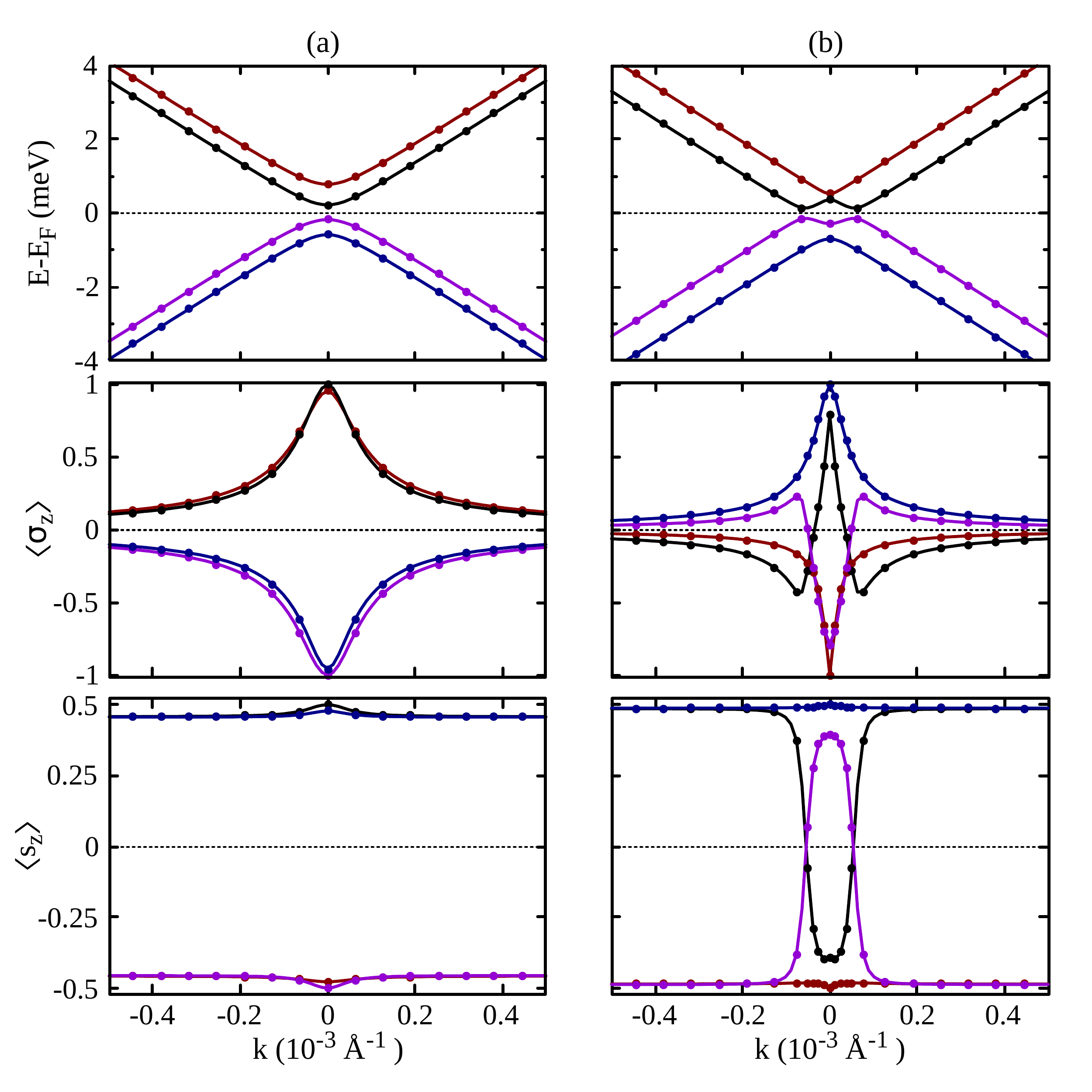}
	\caption{(Color online)  Effective Hamiltonian fitting (solid circle) to tight-binding model (color lines) near K valley. (a) and (b) show two different phases of the system corresponding to Fig.\ \ref{Fig2}, (a) direct band, and (b) inverted band gap phases, respectively. Expectation value of the pseudospin as well as the spin near K for the lowest four bands close to Fermi level are shown in the two lower panels, respectively. Brown, black, purple and blue colors represent corresponding bands of the top panel. The effective Hamiltonian provides an excellent fit to the tight-binding model results \cite{Abdulrhman2016}.}  
	\label{Fig3}
\end{figure}

To gain more insight into this gate dependent phase transition, we study the spin $\langle S_z\rangle$ and AB lattice pseudospin $\langle \sigma_z\rangle$ content of the eigenstates. The bands closest to the neutrality point are characterized using equations
\begin{equation}\label{szgz}
\langle s_z\rangle=\frac{\hbar}{2}\langle\varPsi_i\arrowvert {\sigma}_0\otimes s_z\arrowvert\varPsi_i\rangle, \qquad
\langle\sigma_z\rangle=\langle\varPsi_i\arrowvert\sigma_z\otimes {s}_0\arrowvert\varPsi_i\rangle,
\end{equation}
for each state $\arrowvert\varPsi_i\rangle$. We find that, as shown in lowest panels of Fig.\ \ref{Fig3}, the spin is uniform around the K and K$ ' $ valleys, each band with well-defined spin $\langle  S_z \rangle = \pm\frac{1}{2} \hbar$, which reverses as we cross from K to K$ ' $ valley. This is similar to the behavior of spin states of MoS$_2$ monolayer \cite{mos2}. The pseudospin texture at both valleys is the same, vanishes away from K and K$ ' $, and acquires a value of $\pm1$ as one approaches these points, as shown in the middle panel of Fig.\ \ref{Fig3}.


\subsection{Effective Hamiltonian}
To better understand the physics of this heterostructure system, an effective model that describes the results is vital. Based on the symmetries of the graphene-MoS$_2$ heterostructure, we propose the following Hamiltonian for the states near neutrality, where all the terms respect time reversal symmetry \cite{mahmoud-PRB,mahmoud-PRL,mogrf2}:
\begin{equation}\label{h-eff}
\mathcal{H}_{\rm{eff}}=\mathcal{H}_{\circ}+\mathcal{H}_{\Delta}+\mathcal{H}_{S_1}+\mathcal{H}_{S_2}+\mathcal{H}_R,
\end{equation}
with
\begin{equation}\label{eff-term}
\begin{split}
\mathcal{H}_{0}&=\hbar v_F \left( \tau_z\sigma_{x}p_{x}+\tau_0\sigma_{y}p_{y}\right)s_0\\
\mathcal{H}_{\Delta}&=\Delta s_0 \sigma_z \tau_0\\
\mathcal{H}_{S_1}&=S_1 \tau_z\sigma_zs_z\\
\mathcal{H}_{S_2}&=S_2\tau_z\sigma_0s_z\\
\mathcal{H}_R&= R (\tau_z\sigma_xs_y-\tau_0\sigma_ys_x),
\end{split}
\end{equation}
where $t, \Delta, S_1, S_2$ and $R$ are constants to be found by fitting them to the tight-binding or first principles band structure. $ \sigma_i$, $ \tau_i$, $s_i$ are Pauli matrices where $i={0, x, y, z}$, ( 0 is used for the unit matrix); $\sigma_i$ acts on the pseudospin A, B space, $ \tau_i $ on the K, K$ ' $ valley space, and $ s_i $ operates on the spin degree of freedom. $\mathcal{H}_0$ describes pristine graphene at low energy \cite{qshe}.

Let us analyze the effect of some terms in this effective model. $ \mathcal{H}_{S_2}$ (diagonal spin orbit coupling) breaks the particle-hole symmetry by oppositely shifting bands. The staggered potential ($ \mathcal{H}_{\Delta}$) opens a gap in the otherwise linear dispersion of $\mathcal{H}_{0}$ \cite{mahmoud-PRB,mahmoud-PRL}, and characterizes an asymmetry in the atoms at A and B sublattices. Intrinsic SOC ($\mathcal{H}_{S_1}$) also opens a gap in the bulk structure but with opposite signs at K and K$'$ valleys. Finally, due to the existence of a substrate, mirror symmetry is broken, allowing for we introduce a Rashba effective term ($\mathcal{H}_R$) \cite{mahmoud-PRB, qshe}. The basis of this Hamiltonian is $\Psi_k^T= (A\uparrow,B\uparrow,A\downarrow,B\downarrow)$. We analytically find the parameters at zero momentum $\boldsymbol{k}=0$ that both fit the band structure and satisfy the spin and pseudospin expectation values in Eq.\ \eqref{szgz} for different gate voltages. 

Top panel of Fig.~\ref{Fig3} shows that the effective model fits well with the tight-binding results in Fig.\ \ref{Fig2}. It also captures the essential characteristics of the eigenvectors, including spin and pseudospin expectation values, middle and bottom panels of Fig.\ \ref{Fig3}, respectively. Analyzing these eigenfunctions, we find that the states for conduction bands at the K point reside on the A sublattice, while the valence band states are located on the B sublattice. For the K$ ' $ valley it is the same, with reversed spins.

The parameters of the effective Hamiltonian change smoothly with gate voltage, as shown in Fig.\ \ref{Fig8}(b). In the inverted band phase, the absolute value of the spin orbit interactions $\lvert S_2\rvert$ are enhanced as the gate voltage shifts the Dirac point closer to the valence bands of MoS$_2$, while the staggered interaction $\lvert \Delta\rvert$ term value is smaller than the spin-orbit amplitudes $|S_2|$. 

At the semi metallic phase, both terms are nearly equal, while for trivial gap band gate voltages, the staggered term $\lvert \Delta\rvert$ overcomes the diagonal spin value $\lvert S_2\rvert$. Notice that large diagonal spin orbit coupling and staggered potential terms that characterize the dynamics near K and K$'$ valleys are very similar to the structure in TMDs.
Away from the K point the heterostructure exhibits linear dispersion with a very slight drop in Fermi velocity ($\simeq$ 2\%), nearly independent of V$_g$.

\section{Berry curvature and Chern number}

\begin{figure}
	\includegraphics[width=11.5cm, keepaspectratio=true]{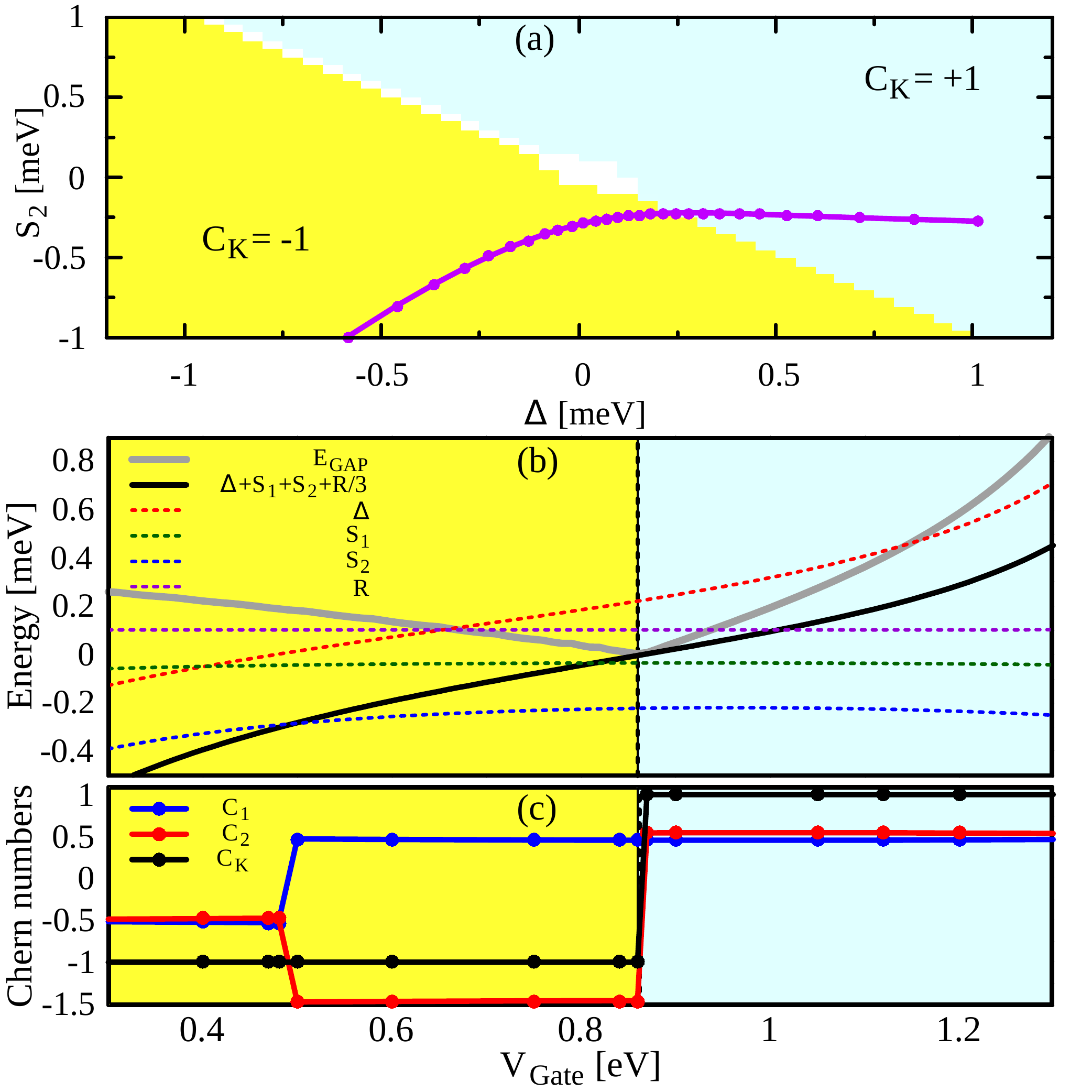}
	\caption{(Color online) 
		(a) 
		Phase diagram of graphene/TMD system in Eq. \ref{eff-term} in the $S_{2}$-$\Delta$ plane with $R$ = 0.1 meV, and $S_{1}$ = $-0.16$ meV. Trivial insulating phase in blue $C_{K}=1$, and mass inverted phase in yellow $C_{K}=-1$, are divided by the semimetallic phase, white curve. Purple line shows the line cut for graphene/MoS$_{2}$ bilayer system as a function of the V$_{Gate}$. 
		(b) Effective parameter dependence on gate voltage corresponding to system in Fig.\ \ref{Fig2}(a). The gap closing occurs at $V_{Gate} = 0.86$ eV, as shown by gray line. 
		Notice that the inverted band regime show a larger SOC contribution, whereas direct band phase show staggered term dominance.
		(c) Corresponding Chern numbers for K valley valence bands (red and blue lines) and total Chern number (black lines) as gate voltage increases.
		Figure \ref{Fig5}(a) explains change in the Chern number near $V_{Gate}=0.5$ eV as due to a band crossing at the K point. The jump at  $V_{Gate}=0.86$ eV indicates gap closing that separates inverted mass regime from direct band regime. \cite{Abdulrhman2016}.}
	\label{Fig8}
\end{figure}

The effective Hamiltonian provides a reliable description of the graphene-MoS$_2$ heterostructure at different gate voltages. It also allows us to further investigate the topology of bands heterostructure. The nature of the gapped phases generated due to the application of an effective gate voltage can be characterized by calculating Berry curvature  $\Omega_{n}(\boldsymbol{k})$ and Chern number per valley $\mathcal{C}_n$ \cite{berry} for the valence bands nearest the gap band using the following equations: 
\begin{equation}\label{berry}
\centering
\begin{split}
\Omega_{n}(\boldsymbol{k})=-\sum_{n'\neq n}\frac{2\rm{Im}\langle\Psi_{n'\boldsymbol{k}}|{v}_x|\Psi_{n\boldsymbol{k}}\rangle\langle\Psi_{n\boldsymbol{k}}|{v}_y|\Psi_{n'\boldsymbol{k}}\rangle}{(\epsilon_n-\epsilon_{n'})^2},
\end{split}
\end{equation}
\begin{equation}\label{chern}
\centering
\begin{split}
\mathcal{C}_n=\frac{1}{2\pi}\int dk_x dk_y \text{ }\Omega_n(k_x,k_y),
\end{split}
\end{equation}
where n is the band number, ${\rm v_x}$(${\rm v_y}$) is the velocity operator along the x(y) direction \cite{graphene-Chern}. Notice that the total Chern number or Berry curvature per valley has the contribution from both valence bands in each valley, e.g., is $\mathcal{C}_{\tau}=\sum_{n=1,2}\mathcal{C}_n$, with $\tau=K$ or $K'$, as appropriate.

\begin{figure}[hb!]
	\includegraphics[width=11.5cm, keepaspectratio=true]{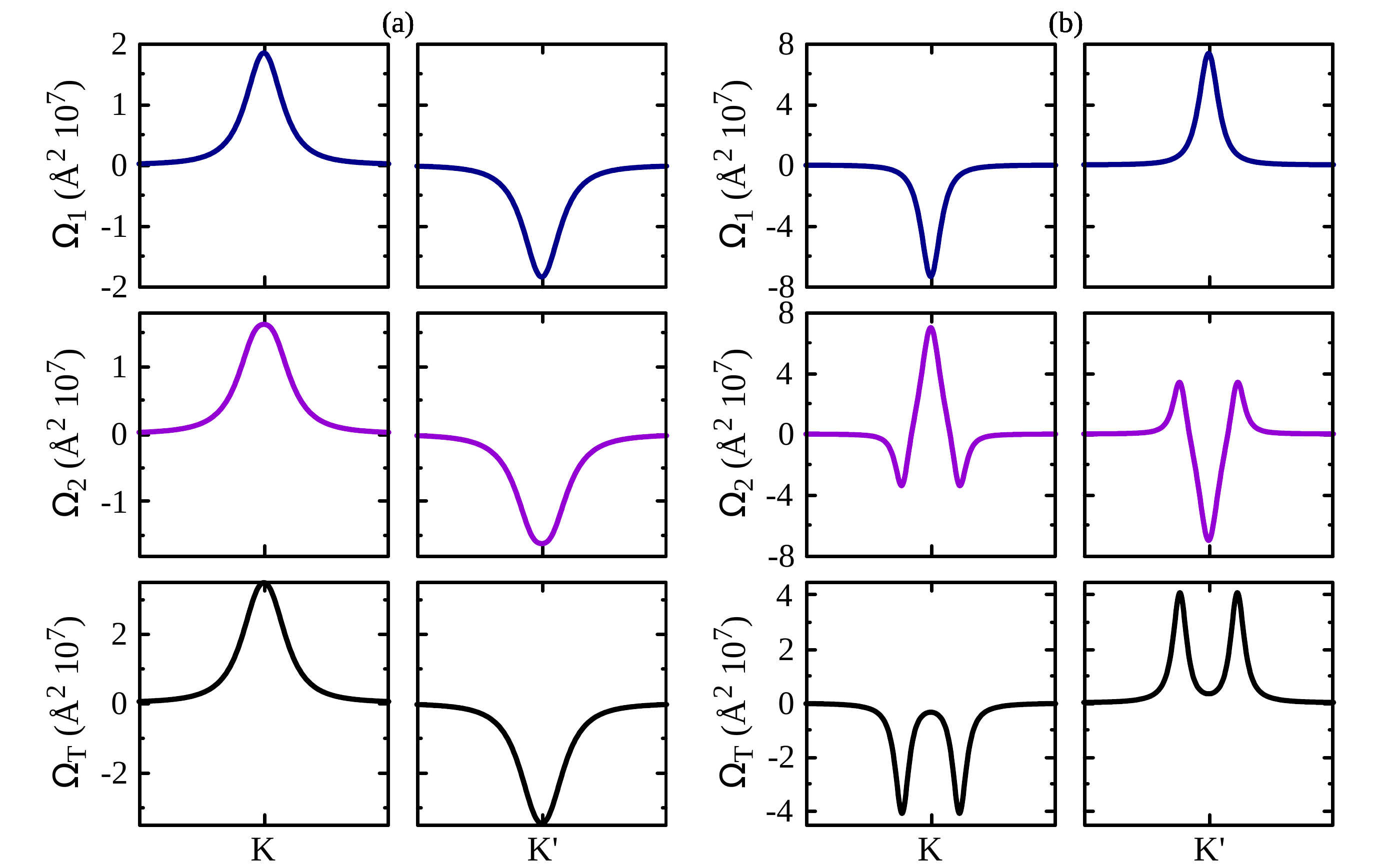}
	\caption{Berry curvature $\Omega_n$ at K and K$'$ valleys for both inverted and direct band gap regimes. (a) Left two columns show results
		for the direct band regime corresponding to Fig.\ \ref{Fig2}(a).  Right two columns are for the inverted band regime corresponding to Fig.\ \ref{Fig2}(b).
		Upper (middle) plots describe Berry curvature of the lowest (highest) energy valence bands in Fig.\ref{Fig3}, $n=1(2)$. Lower plots show
		the total valence band Berry curvature, $\Omega_T = \Omega_1 + \Omega_2$. Time reversal symmetry dictates that Berry curvature in K is reversed in K$'$ \cite{Abdulrhman2016}. }
	\label{Fig6}
\end{figure}

\begin{figure}[hb!]
	\includegraphics[width=11.5cm, keepaspectratio=true]{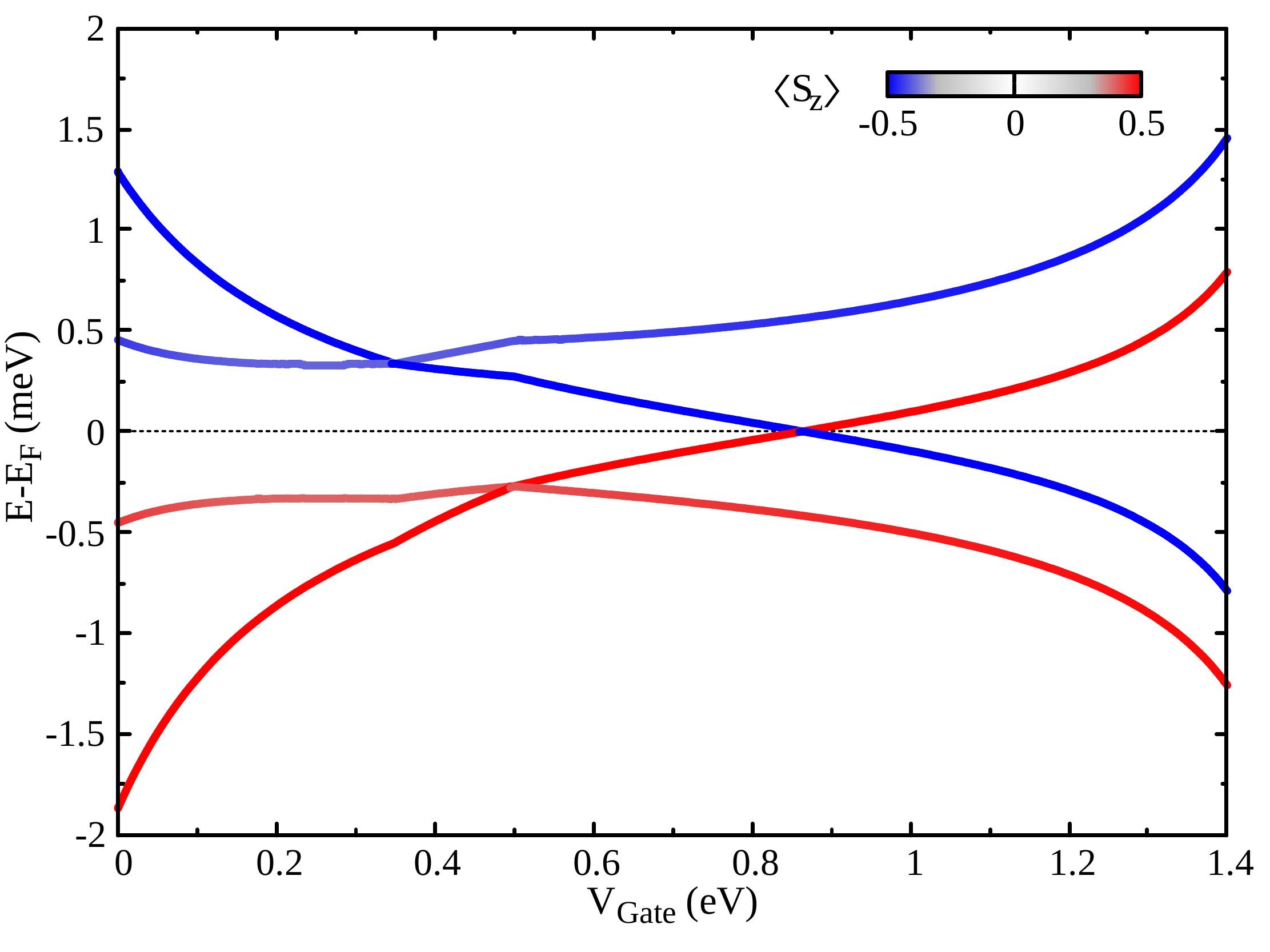}
	\caption{(Color online) G-TMD heterostructure energy values of the four bands around Fermi level at K valley ($ \boldsymbol{k} $ = 0) as a function of gate voltage V$_{\rm Gate}$. Inverted band gap regime is at voltages interval V$_{\rm Gate}\approxeq[0,0.86)$, whereas direct band gap regime is at voltages interval V$_{\rm Gate}\approxeq(0.86,1.4)$. Notice that crossing of bands at the Fermi level at V$_{\rm Gate}\approxeq0.91$ is a semi-metallic state where bulk band gap close. Blue (red) colors describe spin up (down) states \cite{Abdulrhman2016}.
	}  
	\label{Fig5}
\end{figure}

Figure \ref{Fig6} shows the Berry curvature for each valence band and the total curvature per valley around both the K and K$'$ valleys for two regimes: inverted and direct bandgap phases.  We notice that, contrary to the direct band which shows the same curvature for both bands in each valley, the inverted band regime exhibits distinct non-monotonic $k$-dependent curvature in each valley.  

It has been shown that interesting edge states in systems with borders are accompanied by a non-vanishing Berry curvature in each valley, as seen in graphene ribbons and TMD flake edges \cite{ribons2,Li2011,CarlosEdges}. Notice that time reversal symmetry dictates that $\Omega$(K-valley$) = -\Omega$(K$ ' $-valley), as seen in Fig.\ \ref{Fig6}. \cite{berry}

Similarly, a system that preserve time reversal symmetry yields a zero total Chern number,  $\mathcal{C}_K=-\mathcal{C}_{K'}$ \cite{theory,berry}.
However, details of the Chern number per valley differ depending on the topological phase of the system. The sum of the Chern number per valley is non-zero at both the direct bandgap regime and the inverted bandgap regime but with two different values $\pm$1. The sign change at the semimetallic phase predicts a topological phase transition in this system [see Fig.\ \ref{Fig8}(b)]. The competition between the coupling parameters, {\it i.e.} staggered potential and SOC, controls the topological phase of this system, as seen in Fig.\ \ref{Fig8}(a) and \ref{Fig8}(b).

Finally, TMDs substrates play a major role in determining the magnitude of effective parameters induced onto the graphene layer due to proximity effects. The most important of these parameters are the diagonal SOC and the staggered potential. In Fig.\ \ref{Fig8}(a), we plot the Chern number per valley phase diagram by varying these two parameters. We see that, a topological phase transition between the mass-inverted and the direct bandgap regimes is a generic feature, which is expected to exist for all TMD substrates.

\section{Conclusions}
Superimposing graphene and two-dimensional TMDs leads to the appearance of a moir\'{e} pattern which produces a number of interesting effects in these hybrid heterostructures.  The interaction between the layers results in an effective lattice symmetry breaking for the low-energy graphene-like states of the structure.  The states near Dirac points of graphene experience sizable sublattice asymmetry and spin-orbit coupling due to the proximity of the TMD layer, opening a gap in the band structure, akin to the effect of boron nitride on graphene.  However, the strong spin-orbit coupling in TMDs transfers to the graphene states with dramatic consequences.  It is important to note that the relatively weak van der Waals interactions between layers leave most of the linear-dispersion (Dirac cone) of graphene intact, and no charge transfer occurs between graphene and MoS$_2$ layers at ambient conditions. However, the reduced symmetries break the Dirac point singularity and opens bandgap of few meV in the electronic bandstructure, with eigenstates having subtle spin and sublattice spinor content. 

One can moreover realize unique control over the charge-transfer phenomenon between the MoS$_2$ and graphene layers by means of strain or applied gate voltages between the layers. We have shown that the direct bandgap can be significantly tuned by applying biaxial strain on the MoS$_2$ substrate. We have further analyzed that an interlayer effective gate voltage can drive the system through a phase transition between a trivial direct band and a non-trivial inverted bandgap structure. The latter phase is achieved whenever the neutrality point is shifted towards the valence bands of the TMD, as the spin-orbit coupling is found to dominate over the sublattice asymmetry (staggered) effect in that regime.  

The agreement between first-principles and tight-binding model calculations assures that this predicted effect is robust and should be observable in experiments.  Moreover, the effective Hamiltonian is suggestive that this behavior is quite general, and that other TMDs would have similar effects on graphene, as characterized by complex Berry curvature and corresponding Chern numbers. The notion of being able to drive a material system across a topological transition is interesting.  However, the possibilities of achieving quantum spin Hall and valley Hall effects in such system, when a finite-size structure is driven into the inverted bandgap regime are indeed tantalizing.

\section{Future Directions}

The growing interest on hybrid heterostructures of graphene/MoS$_2$, in general graphene/TMD or graphene/2D-material for new physical behavior or technological applications calls for more theoretical research in this direction. Researchers have started exploring the effect of intercalation of small metallic ions between the two monolayers of graphene and TMD. It would be interesting to study the diffusion, adsorption, and intercalation of various different types of elements on the electronic, mechanical, thermal, photoluminescence, energy storage, and catalytic properties of such heterostructures. Changes in the properties due to substrate-induced or externally-applied strain, as well as the formation of domain walls or grain boundaries, is a subject to investigate further--especially as large area samples may exhibit multiple domains. 

The presence of distinct topological phases in graphene/TMD heterostructures indicates there is a need to characterize distinct topological states using topological invariant numbers, as done for topological insulators. A Weyl semimetallic phase may be realized at the trivial to non-trivial topological insulators phase boundary in similar heterostructures. The proximity effect of physisorption of another graphene monolayer or another MoS$_2$ monolayer on graphene/TMD heterostructures might yield intriguing phenomena in the electronic and thermal properties of these heterostructures. The effects due to doping, vacancies, chemical pressure, temperature, and anisotropic strain require systematic studies for the thorough exploration of these systems before devices can be fully developed. Lastly, since these heterostructures seem to be promising candidates for bio-sensing applications, changes in their properties in contact with chemical or biological entities would be important to pursue.

\begin{acknowledgement}

This work used the Extreme Science and Engineering Discovery Environment (XSEDE), which is supported by National Science Foundation grant number OCI-1053575. Additionally, the authors acknowledge the support from Texas Advances Computer Center (TACC), Bridges supercomputer at Pittsburgh Supercomputer Center and Super Computing Systems (Spruce and Mountaineer) at West Virginia University (WVU). A.H.R. and S.S. acknowledge the support from National Science Foundation (NSF) DMREF-NSF 1434897 and DOE DE-SC0016176 projects. S.S. also acknowledges the support from the Robert T. Bruhn research award, Dr. Mohindar S. Seehra Research Award, and the WVU Foundation Distinguished Doctoral Fellowship. S.E.U. and A.M.A. acknowledge the support from NSF DMR 1508325 (Ohio), and the Saudi Arabian Cultural Mission to USA for a Graduate Scholarship.

\end{acknowledgement}

\section*{Appendix}
\addcontentsline{toc}{section}{Computational Details}

\section{Computational Details}
We use the projector augmented-wave (PAW) method as implemented in the {\sc VASP} code \cite{Kresse1996, Kresse1999} to carry out all Density Functional Theory (DFT) \cite{HK1964, KS1965} based first-principles calculations reported in this chapter. Perdew-Burke-Ernzerhof (PBE) parametrized generalized gradient approximation (GGA) was employed for exchange-correlation functional. \cite{Perdew1996}  Twelve valence electrons of Mo (4$p^6$, 5$s^1$, 4$d^5$), six valence electrons of S (3$s^2$, 3$p^4$), and four valence electrons of C (2$s^2$, 2$p^2$) were considered in the PAW pseudo-potential. In order to minimize the lattice mismatch between graphene and MoS$_2$ layers, we consider following two supercell geometries to construct the graphene/MoS$_2$ bilayer heterostructure: (i) 5:4 and (ii) 4:3. A vacuum thicker than 17 \AA~was added along $c$-axis to ensure no interaction between two periodically repeated cells along c-axis. The lattice parameters and the inner coordinates of atoms were optimized until the total residual forces were less than $10^{-4}$ eV/{\AA} per atom, and $10^{-8}$ eV was defined as the total energy difference criterion for convergence of electronic self-consistent calculations. Spin-orbit coupling (SOC) and van der Waals (vdW) interactions were included in the structural optimization. The Tkatchenko-Scheffler (TS) method \cite{TS_PRL2009, Bu_TSvdW2013} was employed for the non-local vdW corrections in the DFT calculations. $650$ eV was used as the kinetic energy cutoff of plane wave basis set and a $\Gamma$-type $10 \times 10 \times 1$ $k$-point used to sample the irreducible Brillouin zone. To investigate the effect of in-plane strain on Mo-atoms, we strained Mo-atoms within the optimized unit cell of 5:4 bilayer heterostructures, while performing full relaxation of the S atoms. The {\sc PyProcar} code \cite{PyProcar, sobhit2016PRB} was used to plot the spin-projected electronic bands and {\sc VESTA} software \cite{VESTA} was used to make figures for the crystal structure and plot the isosurface charge density.

 \bibliographystyle{ieeetr}
  \bibliography{cites}

\end{document}